\documentclass[aps,prd,reprint,superscriptaddress,nofootinbib]{revtex4-2}
\usepackage[english]{babel}
\usepackage[utf8]{inputenc}

\usepackage[pdftex]{hyperref} 
\usepackage{comment}
\usepackage{graphicx}
\usepackage{color}
\usepackage{natbib}
\usepackage{float}
\usepackage{xcolor}
\usepackage{amsmath,amssymb,amsfonts}
\usepackage[capitalise]{cleveref}
\usepackage{subfigure}
\usepackage{orcidlink}
\usepackage{soul}

\newcommand{\mnras}{Mon.~Not.~R.~Astron.~Soc.}
\newcommand{\apjl}{Astrophys.~J.~Lett.}
\newcommand{\aap}{Astron.~Astrophys.}

\newcommand{\zphya}{Z.~Phys.~A}
\newcommand{\natast}{Nat.~Astron.}

\newcommand{\universe}{Universe}

\newcommand{\physrep}{Phys.~Rep.}

\newcommand{\sovast}{Sov.~Astron.}
\newcommand{\epja}{Eur.~Phys.~J.~A}
\newcommand{\science}{Science}
\newcommand{\apjs}{Astrophys.~J.~Suppl.}
\newcommand{\arnps}{Annu.~Rev.~Nucl.~Part.~Sci.}
\newcommand{\ptep}{Prog.~Theor.~Exp.~Phys.} %Progress of Theoretical and Experimental Physics

\newcommand{\nucphysa}{Nucl.~Phys.~A}
\newcommand{\nucphys}{Nucl.~Phys.}

\newcommand{\ptp}{Prog.~Theor.~Phys.}
\newcommand{\physletb}{Phys.~Lett.~B}
\newcommand{\plb}{Phys.~Lett.~B}
\newcommand{\ppnp}{Prog.~Part.~Nucl.~Phys.}

\newcommand{\pasa}{Publ.~Astron.~Soc.~Aust.}
\newcommand{\jpg}{J.~Phys.~G}
\newcommand{\ppn}{Phys.~Part.~Nucl.} %Physics of Particles and Nuclei

\newcommand{\epjstiso}{Eur.~Phys.~J.~Spec.~Top.}
\newcommand{\apss}{Astrophys.~Space~Sci.} % Astrophysics and Space Science
\newcommand{\physscr}{Phys.~Scr.} % Physica Scripta

\addto\captionsenglish{}

\begin{document}

\title{
Delayed collapse of massive proto-neutron stars to a black hole as a signature of non-nucleonic degrees of freedom
}

\author{Hristijan Kochankovski}
\email[Corresponding author: ] {hriskoch@fqa.ub.edu} 
\affiliation{Departament de F\'isica Qu\`antica i Astrof\'isica and Institut de Ci\`encies del Cosmos (ICCUB), Facultat de F\'isica, Universitat de Barcelona, Mart\'i i Franqu\`es 1, 08028 Barcelona, Spain}

\author{Ishfaq Ahmad Rather~\orcidlink{0000-0001-5930-7179}}
\email{rather@astro.uni-frankfurt.de}
\affiliation{Institut f\"{u}r Theoretische Physik, Goethe Universit\"{a}t,
Max-von-Laue-Str.~1, D-60438 Frankfurt am Main, Germany}

\author{Jan-Erik Christian}
\email{jan-erik.christian@uni-hamburg.de}
\affiliation{Hamburger Sternwarte, University of Hamburg, Gojenbergsweg 112, 21029 Hamburg, Germany}

\author{\`Angels Ramos}
\email{ramos@fqa.ub.edu} 
\affiliation{Departament de F\'isica Qu\`antica i Astrof\'isica and Institut de Ci\`encies del Cosmos (ICCUB), Facultat de F\'isica, Universitat de Barcelona, Mart\'i i Franqu\`es 1, 08028 Barcelona, Spain}

\author{Jürgen Schaffner-Bielich~\orcidlink{0000-0002-0079-6841}}
\affiliation{Institut f\"{u}r Theoretische Physik, Goethe Universit\"{a}t,
Max-von-Laue-Str.~1, D-60438 Frankfurt am Main, Germany}

\author{Laura Tolos~\orcidlink{0000-0003-2304-7496}}
 \email{tolos@ice.csic.es}
\affiliation{Institute of Space Sciences (ICE, CSIC), Campus UAB, Carrer de Can Magrans, 08193 Barcelona, Spain}
\affiliation{Institut d'Estudis Espacials de Catalunya (IEEC), 08860 Castelldefels (Barcelona), Spain}

\date{\today} 

\begin{abstract}

We study the stability of the massive proto-neutron stars (PNSs) against a delayed collapse to a black hole during their early evolution at constant baryonic mass. We employ a large and heterogeneous set of publicly available finite-temperature equations of state (EoSs), 16 of them containing additional hadrons besides nucleons (hyperons, $\Delta$ resonances and (anti)kaon condensates) and 23 purely nucleonic ones. We find that 14 of the 16 EoSs with additional hadronic degrees of freedom predict a delayed collapse of the most massive PNSs, the two exceptions having a very scarce hyperon population, whereas none of the nucleonic EoSs do. By separating the effects of deleptonization from those of heating and cooling, we show that the nucleonic and hyperonic families differ in their response to deleptonization, which softens the EoS when hyperons are present and stiffens it otherwise. Heating lowers the maximum baryonic mass of both families, slightly more in the presence of non-nucleonic hadrons, and for some models it is what finally triggers the collapse. We moreover study 39 additional relativistic mean-field nucleonic parametrizations compatible with chiral effective field theory constraints on neutron matter, spanning a wide range of effective nucleon masses and symmetry energy parameters. We find that none of them predict a delayed collapse. Hence, our results identify delayed collapse in the most massive PNSs as a robust fingerprint of non-nucleonic hadronic degrees of freedom in hot, dense matter, provided they appear in sizable amounts.

\end{abstract}


\maketitle

\section{Introduction} \label{sec:introduction}

Neutron stars (NSs) are among the densest objects in the Universe and serve as unique laboratories for probing the behavior of matter at extreme conditions. Beyond nucleon-lepton matter, exotic matter may appear as additional degrees of freedom in their cores, where the density is well above nuclear saturation density ($\rho_0\simeq 0.16\,\mathrm{fm^{-3}}$)~\cite{SchaffnerBielichCompactStarPhysicsbook2020, Tolos:2020aln, Vidana:2018bdi}. In particular, hyperons, baryons containing strange quarks, are among the most probable candidates to be present in the inner core of NSs~\cite{AmbartsumyanHyperons1960, GlendenningHyperonCompositionNS1982, GlendenningNSGiantYNuclei1985, GlendenningHyperonsNS1987, GlendenningNSMasses1991, KnrrenQuarkHadronPhaseTransitionNS1995, BalbergEffectiveEoSStrangeness1997, SchulzeYNMatterBHF1998, BaldoOnsetYNSBrueckner1998, BalbergRolesYNS1999, BaldoYStarsBBG2000, VidanaStrangeNuclearMatterBHF2000, VidanaYYInteractionPropertiesNSMatter2000, Schaffner:1995th, Schaffner-Bielich:2000nft, Providencia:2012rx}. Their presence can have a sizable impact on the equation of state (EoS), which is controlled by the poorly constrained hyperon-nucleon and hyperon-hyperon couplings whose variation within the allowed SU(3) and hypernuclear range changes the astrophysical observables of NSs~\cite{Weissenborn:2011kb, Weissenborn:2011ut, Fortin:2017cvt}.  
Besides hyperons, $\Delta$ resonances and an antikaon condensate may become energetically favorable at these densities. In the case of the antikaons, the relevant quantity is the $\Bar{K}$ optical potential in dense matter, which is attractive but whose depth is still debated, both in chiral coupled-channel calculations~\cite{Tolos:2002ud, Tolos:2006ny, Hrtankova:2019jky, Bruns:2021krp, Cieply:2023saa, Obertova:2025rso} and in the analysis of kaonic atoms~\cite{Friedman:2007zza, Feijoo:2026eof, Yamagata-Sekihara:2024zrk}, and which benefits from two- and three-body constraints provided by femtoscopic measurements of the $K^- p$ and $K^- d$ interactions~\cite{ALICE:2021szj,ALICE:2026pxr,Ramos:2025ibe}, although extrapolating these to high baryon densities still requires many-body modeling. For the $\Delta$ resonances, the main source of uncertainty is their interaction with the nuclear medium, which is only loosely constrained by experimental data, and a moderately attractive $\Delta$ potential is enough to bring the onset of the $\Delta^-$ close to, or even below, that of the first hyperons~\cite{Drago:2014oja, Li:2018qaw, RadutaEoSHotNSExoticParticleDoF2022,Ribes:2019kno}.

In the last two decades, considerable progress has been achieved in constraining the properties of cold NSs from astrophysical observations. Precise mass measurements of massive pulsars with $M \gtrsim 2\,M_{\odot}$~\cite{DemorestSharipoDelay2MNS2010, AntoniadisMassivePulsarCompactBinary2013, FonsecaNANOGRAVBinaryMilisecondPulsar2016, CromartieRelativisticSharipoMAssiveMilisecondPulsar2020, RomaniFastestheaviesNS2022} have placed stringent constraints on the stiffness of the EoS at high densities. Furthermore, the X-ray timing observations from the NICER mission have provided simultaneous mass-radius constraints~\cite{RileyNICERViewMilisecondPulsar2019, MillerMassRadiusNICERData2019, RileyNicerViewRadioTiming2021, MillerRadiusFromNICER2021, VinciguerraUpdatedMassRadiusAnalysisNICER2024, MauviardANicerView14MSunPulsar2025, ChoudhuryNICERViewNearestBrightestMillisecondPulsarJ04372024, SalmiRadiusHighMassPulsarJ0740NICERData2024, MauviardNICERViewJ1614MassiveCompactMillisecondPulsar2026, Stammler:2026xaq}, adding valuable information on the dense matter EoS. However, these observables are related to cold, isolated NSs and, therefore, to the one-dimensional, barotropic EoS. They are also largely blind to the composition, as a model containing hyperons, or any other non-nucleonic species, can reproduce a mass-radius relation that is nearly indistinguishable from a purely nucleonic one. This issue, equivalent to the ``masquerade problem'' which was raised in the context of the hadron-quark phase transition~\cite{AlfordHybridStarsMasqueradeNS2005}, implies that, as long as the interactions and the EoS of purely nucleonic matter are not fully settled, it may be difficult to unambiguously determine if hyperons occur in NSs from cold observables alone. Nevertheless, promising paths to disentangle the effects of the composition on cold NS observables have been sketched recently (see, for example, Refs.~\cite{FerreiraIdentifyingHyperonsNSMRSlope2025, BausweinStellarPropertiesIndicatingY2025, Carvalho:2024bxv, Malik:2022jqc, Sanson:2026lky}). 
A degeneracy of the same kind affects the identification of a strong first-order transition, where very similar mass-radius and tidal deformability data are reproduced by models with and without a disconnected branch~\cite{Christian:2018jyd, Christian:2021uhd}. A complementary route is the oscillation modes of the star, which respond to the composition in a way that the static mass-radius relation does not: the additional degrees of freedom shift the fundamental ($f$-mode) frequency and, more specifically, the composition gradients they introduce support buoyancy-driven $g$-modes that are absent in purely nucleonic, barotropic matter~\cite{Rather:2024mtd, BarmanInvestigatingRoleNuclearParametersOscillationModesHotNS2025, Thakur:2025qwl, Thakur:2026cac}. 

Contrary to cold, already evolved NSs, the early stages of the NS life probe the generalized three-dimensional EoS, which, in addition to density, also depends on temperature and lepton fraction. Right after the core collapse of a massive star, the newly born proto-neutron star (PNS) is a hot, lepton-rich object, where neutrinos are initially trapped within the dense medium. As the PNS evolves through the Kelvin-Helmholtz cooling phase, neutrinos gradually diffuse out; the star deleptonizes and cools down to form a cold, catalyzed NS.  In many recent works, the influence of the exotic baryons on the thermal properties of the EoS has been studied in detail~\cite{KochankovskiEoSHotHyperonicNSM2022, KochankovskiHyperonicUncertaintiesNS2024, RadutaPNSHeavyBaryons2020, RadutaEoSHotNSExoticParticleDoF2022, SedrakianEoSPNSMergerRemnantsHyperons2021, SedrakianDeltaResonancesHyperonsPNSMergerRemnants2022, GuichonHFQMCFiniteT2024, MalikEoSHyperonsKaonsSupernovaNSMergers2021, StoneEoSHotYMatterQMCA2021, TsiopelasFiniteTEoSCompactStarsY2024}. Temperature affects the EoS in two distinct ways. On the one hand, it provides a direct thermal contribution to the pressure at fixed composition. On the other hand, it modifies the composition itself: thermal excitations populate hyperons at densities below their cold threshold and enhance their abundance above it,  converting nucleons near the top of the Fermi sea into heavier baryons that contribute less pressure at the same baryon density, so that the EoS softens. In purely nucleonic matter, no new species become available, and only the first effect operates, whereas in the presence of exotic degrees of freedom the two compete. The different composition of matter at finite temperature is precisely what makes the thermal behavior of the EoS a potential discriminator between hyperonic and purely nucleonic matter~\cite{BlackerThermalBehaviorIndicatorHyperonsBNSMergers2024, KochankovskiEoSHotHyperonicNSM2022, RadutaEoSHotNSExoticParticleDoF2022}. 
Any such statement, however, has to be compared against the thermal behavior of purely nucleonic matter as it is predicted by ab initio calculations, which recently have been constrained directly from chiral effective field theory interactions at finite temperature~\cite{Carbone:2019pkr, Keller:2020qhx, Keller:2022crb}.
The most massive stars are especially interesting in this context, as the amount of exotic components, if present, at the high densities reached in their cores should be the highest.

A possible clear signal of the appearance of exotic hadrons in the early evolution of a NS was put forward already three decades ago~\cite{BrownBetheScenarioLowMassBHGalaxy1994, ThorssonCompositionStructureEvolutionNSKaonCondensates1994, KeilJankaHadronicPhaseTransitionsDelayedCollapse1995, GlendenningPromptSubsidencePNSBlackHole1995, BombaciMmaxNS1996}.  
 It was shown that negatively charged hadrons, such as hyperons and kaons, may cause an initially stable PNS to become unstable during its subsequent cooling, resulting in a delayed collapse of a PNS to a black hole. Although baryonic mass is conserved during cooling in the absence of accretion, stellar deleptonization increases the population of negatively charged hadrons, softening the EoS. The maximum gravitational-mass configuration at that cooling stage, governed by the EoS, would correspond to a lower baryonic mass than the initial maximum-mass PNS, delaying the collapse of the metastable PNS during the first tens of seconds of its life. This scenario was explicitly connected with SN1987A~\cite{BrownBetheScenarioLowMassBHGalaxy1994}, where the absence of a detected pulsar could be explained by such a delayed collapse.\footnote{Although pulsations have still not been observed, recent ALMA and JWST observations provide indirect evidence for a neutron star in the remnant~\cite{CiganHighAngularResolutionALMADustMoleculesSN1987A2019, PageNS1987AinSN1987A2020, FranssonEmissionLinesIonizingRadiationCompactObjectSN1987A2024, LarssonCompactObjectInnermostEjectaSN1987A2025}, which disfavors the interpretation of delayed collapse for SN1987A.}

In the thirty years after the publication of these seminal works, the question of the evolution of the maximum mass of PNSs in the presence of exotic components has been addressed by many different authors. Shortly after the original works, the metastability window created by strangeness was further quantified, and was explored in detail, both in phenomenological and in microscopic approaches~\cite{ PrakashCompositionStructurePNS1997, PonsEvolutionPNS1999, StrobelMinimumMaximumMassNSDelayedCollapse2001, VidanaMicroscopicStudyNeutrinoTrappingHyperonStars2003, YuanRotationalEvolutionPNSHyperons2005, BurgioMicroscopicEoSPNS2007, BurgioHyperonStarsFiniteTBruecknerTheory2011, BanikProbingMetastabiliyPNS2014, FortinHyperonsHotDenseMatter2018, RadutaPNSHeavyBaryons2020, KhadkikarMaximumMassCompactStarsGWFiniteTemperatureEoS2021, FischerHyperonsPNSDeleptonizationDarkFlavouredParticles2025, CanullanPascualNSStabilityBeyondMassPeakOutEquilibriumPerturbations2025, KumarConstraintsMaximumNSPNSEvolution2025, Gulminelli:2013qr,BanikAnitKaonCondensationMetastabilityPNS2001,PonsEvolutionPNSKaonCondensates2001}. This question was also addressed for an antikaon condensate, where the metastability window was found to be comparably wide~\cite{BanikAnitKaonCondensationMetastabilityPNS2001, PonsEvolutionPNSKaonCondensates2001}. The same criterion has recently been applied to color-superconducting quark matter, where a delayed collapse is found among the possible evolutionary outcomes~\cite{Kunkel:2026wvz}. Delayed collapse has also been found for proto-neutron stars with a quark-hadron mixed-phase core \cite{Pons:2001ar}. In a quasi-static study with a constant-speed-of-sound quark phase, collapses occur only rarely, namely when no stable hybrid configuration exists and the transition point is reached \cite{Milthaler:2026dhm}. Dynamical simulations show that a phase-transition-induced collapse can be stopped by a stable hybrid star configuration \cite{Naseri:2024rby}. The outcome therefore depends on the quark-matter model and on how the transition is treated.  It is worth noting that most of the works mentioned above have focused on one or a few EoS models at a time, and a systematic analysis of the PNS maximum mass evolution using a large and diverse set of finite-temperature EoSs is still lacking.

In this article, we revisit the question of the PNS evolution in the presence of exotic hadrons from a different perspective. Instead of studying only a few EoSs, we employ a large and heterogeneous set of publicly available finite-temperature EoSs, which comprises 16 models with additional non-nucleonic hadronic degrees of freedom and 23 purely nucleonic ones, as described in \cref{sec:EoS}. In \cref{sec:PNSevolution}, for each of them we compute the baryonic mass at the maximum gravitational mass star configuration (maximum baryonic mass) at specific snapshots of the PNS evolution, and we compute the difference with respect to the baryonic mass of the initial PNS. A negative value, indicating a delayed collapse, is found for 14 out of the 16 EoSs with additional hadronic degrees of freedom, while none of the 23 purely nucleonic ones do. In \cref{sec:cooling_deleptonization} we show the results for the maximum mass evolution and disentangle the contribution of the deleptonization from that of the heating and cooling of matter, in order to identify which of the two drives the instability. We find that the instability is driven mainly by the deleptonization: at fixed entropy, the loss of leptons lowers the maximum baryonic mass of the models with non-nucleonic degrees of freedom by up to $0.13\,M_{\odot}$, while it raises it by $0.07$ to $0.31\,M_{\odot}$ for every purely nucleonic model. In \cref{sec:other_results} we examine how the non-nucleonic hadronic components modify the structure and the thermal properties of the maximum-mass star. In \cref{sec:RMFNucleonic} we test the robustness of the nucleonic results with a set of Relativistic Mean-Field (RMF) parametrizations constrained by chiral effective field theory, and we explicitly show that a delayed collapse is not found anywhere in the explored parameter space. Finally, we summarize our findings and conclude in \cref{sec:conclusions}.

\section{Finite-temperature Equation of State of dense matter}

\label{sec:EoS}

In this section, we give a brief overview of the different finite-temperature EoSs of dense matter that are employed in the present work. All of them are general-purpose EoS tables that are publicly available on the CompOSE database~\cite{TypelCompOSE2015, TypelCompOSEManual2022, DexheimerQuickGuidesCompOSEDataBase2022}, and cover the wide range of densities, temperatures, and charge fractions explored during the PNS evolution. The set of models is intentionally large and heterogeneous, since it contains purely nucleonic models as well as models that include hyperons, $\Delta$ resonances, and (anti)kaons as additional degrees of freedom\footnote{Note that we refer to hyperonic EoSs when considering nucleons and hyperons and, in certain cases, also $\Delta$ baryons or an antikaon condensate on top of the hyperonic content. We will simply use ``nucleonic'' when we refer to systems which contain only nucleons but no heavier baryons or meson condensates.}. In this way, the conclusions drawn in the subsequent sections do not rely on the particularities of a single model or framework, but rather reflect the systematic differences between the models with and without exotic components. All the models are listed in \cref{table:DelayedCollapse}, together with their theoretical framework, their particle content, and the corresponding references. They can be divided into two broad categories, phenomenological and microscopic ones, which we briefly describe in the following. For detailed reviews of the EoS of dense and hot matter, see, for example, Refs.~\cite{Tolos:2020aln, Vidana:2018bdi, OertelEoSSupernovaeCS2017, BurgioNuclearEoSCompactStarsSupernovae2018, ChatterjeeHyperonsExistNS2016, RadutaEoSHotNS2021}.

Most of the phenomenological models belong to the broad category of covariant RMF density functionals, in which the many-body hadron interactions are modeled with different meson-exchange channels at the mean-field level. Depending on how the medium dependence of the interaction is implemented, they are further divided into models with a non-linear dependence on the meson mean fields (NL-RMF) and models with density-dependent couplings (DD-RMF). Within the first subcategory, the STOS and TM1e tables are based on the TM1 parametrization~\cite{SugaharaRMFUnstableNuclei1994}, whereas the couplings of the QMC-RMF family are constrained by chiral effective field theory calculations of pure neutron matter performed with Quantum Monte Carlo techniques. We note that, despite the similar acronym, the latter family is not related to the quark-meson-coupling scheme mentioned below. The remaining NL-RMF models are IUF, a modification of the FSUGold parametrization, and SFHo and SFHx, which were fitted to neutron star mass-radius observations. The SFHo model was further extended with hyperons in the SFHoY model. The set is completed by the hyperonic FSU2H$^{*}$ model, whose hyperon couplings are fitted to hypernuclear data, and by the nucleonic BCPN and hyperonic BCPY models, which are built on parametrizations fitted to the symmetry energy and its slope in pure neutron matter in agreement with chiral effective field theory calculations.

Within the second subcategory, the DD2, DD2Y, BHB$\Lambda \phi$, BHB$\Lambda \phi$K and DD2K models, the last two of which include an antikaon condensate on top of the hyperonic and nucleonic content, respectively, share the same nucleonic interaction and low-density treatment, and they therefore form a controlled set that allows us to disentangle the effects of the different exotic degrees of freedom. In addition, we consider the DD2Y$\Delta$ family whose models differ in the values of the $\Delta$ couplings and, as a consequence, in the onset density of the $\Delta$ resonances, which are assumed to have negligible widths and unmodified vacuum masses. As pointed out in previous works~\cite{RadutaEoSHotNSExoticParticleDoF2022, KochankovskiImpactHyperonsNSMergers2025, Li:2018qaw, Drago:2014oja, Rather:2023dom, Rather:2024mtd}, the appearance of $\Delta$ resonances has an analogous effect on the EoS as that produced by hyperons and, hence, we include this family in the study. The DDLS family explores the uncertainty with respect to the slope of the symmetry energy ($L = 30, 50$ and $70$ MeV), both in its purely nucleonic and hyperonic versions. Two further covariant schemes are represented by one hyperonic model each. The DNS EoS is based on the SU(3) Chiral Mean Field (CMF) model, in which the baryon masses are generated via the coupling to the scalar meson fields and are significantly lowered when chiral symmetry is restored, while QMC-A is built within the Quark-Meson-Coupling (QMC) model, in which the hadronic interactions arise from the coupling of the valence quarks to the meson fields.

The remaining phenomenological models are non-relativistic and are based on Skyrme interactions~\cite{SkyrmeTheNuclearSurface1956, SkyrmeEffectiveNuclearPotential1958}, which describe contact interactions between nucleons at low momenta. Contrary to the RMF models, they do not fulfill the causality constraint by construction at the highest densities found in NSs. We consider the SRO(SLy4) model and the RB family, a recently developed set of EoSs based on extended Brussels-Skyrme interactions that were generated within a Bayesian analysis constrained by ab initio calculations of pure neutron matter. Of the eight members of this family, three (RB1, RB3 and RB4) are excluded from the present analysis because their tabulated density range ends before the maximum-mass configuration is reached. To the best of our knowledge, there are no publicly available finite-temperature Skyrme EoSs that include hyperons as degrees of freedom.

Contrary to the phenomenological models, the microscopic models start from realistic two- and three-baryon interactions that are fitted to experimental scattering data~\cite{BurgioEoSNuclearMatterFiniteNucleiNS2020, BurgioNSNuclearEoS2021, DrischlerEFTHighDensityNuclearEoS2021}. We consider three of them. The LPB EoS is derived within the finite-temperature Brueckner-Hartree-Fock (BHF) approximation starting from nucleon-nucleon interactions obtained from chiral effective field theory, while SRO(APR) and FTNS are based on variational calculations with the Argonne v18 two-nucleon potential~\cite{WiringaAccurateNNPotentialChargeIndependenceBreaking1995} supplemented with the Urbana IX three-nucleon force~\cite{PudlinerQMCCalculationsA6Nuclei1995}. All three microscopic models are purely nucleonic, since no general-purpose finite-temperature tables with hyperons are currently available within these approaches. Therefore, they serve as additional benchmarks against which the models with exotic components can be compared.

\begin{figure*}[t]
    \centering
    \includegraphics[width=0.90\textwidth]{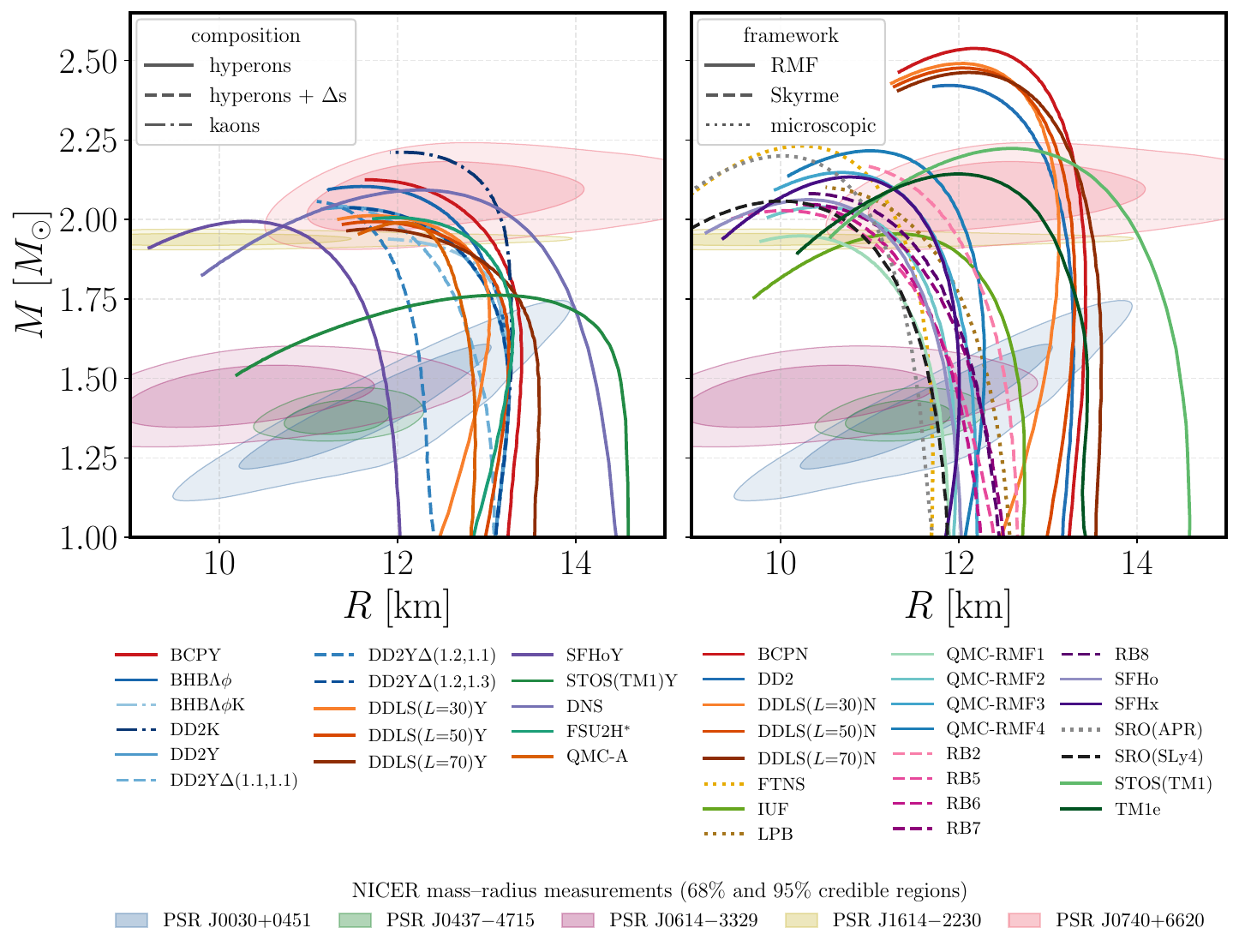}
    \caption{Mass-radius relations of cold, $\beta$-equilibrated NSs for all the EoS models employed in this work. The left panel shows models with hyperons, $\Delta$ resonances, and/or (anti)kaons, with the linestyle indicating the exotic composition; the right panel shows purely nucleonic models, with the linestyle indicating the theoretical framework (RMF, Skyrme, or microscopic). Models built on the same underlying nucleonic interaction are drawn in shades of the same color. The shaded areas show the 68\% and 95\% credible regions of the NICER mass-radius measurements of PSR J0030+0451~\cite{VinciguerraUpdatedMassRadiusAnalysisNICER2024}, PSR J0437$-$4715~\cite{Stammler:2026xaq}, PSR J0614$-$3329~\cite{MauviardANicerView14MSunPulsar2025}, PSR J1614$-$2230~\cite{MauviardNICERViewJ1614MassiveCompactMillisecondPulsar2026} and PSR J0740+6620~\cite{SalmiRadiusHighMassPulsarJ0740NICERData2024}. } 
    \label{fig:cold_MR}
\end{figure*}

The properties of the cold, $\beta$-equilibrated NSs predicted by each model are also collected in \cref{table:DelayedCollapse}, and the corresponding mass-radius curves are shown in \cref{fig:cold_MR}. They serve as an indicator of the wide parameter space covered by our selection of models, which encapsulates the uncertainty present in the effective baryon-baryon interactions. Most of the EoSs are (at least marginally) compatible with the $2\,M_{\odot}$ constraint. However, the BHB$\Lambda\phi$K, DDLS($L$=50)Y, DDLS($L$=70)Y, IUF, QMC-RMF1, QMC-A, SFHoY and, most notably, STOS(TM1)Y models do not satisfy this constraint. We nevertheless include them because the delayed collapse is a property of the internal consistency of each model along its own evolution and not of its agreement with pulsar-mass measurements, so every published table is a legitimate test case. Moreover, the conclusion does not depend on the EoSs that do not satisfy the $2\,M_{\odot}$ constraint: restricting the sample to the models that reach $2\,M_{\odot}$ leaves 10 EoSs with non-nucleonic degrees of freedom, 9 of which collapse, and 21 purely nucleonic ones, none of which does.

\begin{table*}[tp]
\caption{List of all the EoSs used in this work, grouped into purely nucleonic models and models with exotic degrees of freedom. For each model, we give the theoretical framework, the particle content ($N$ for nucleons, $Y$ for hyperons, $\Lambda$ when only $\Lambda$ hyperons are included, $\Delta$ for $\Delta$ resonances, and $K$ for an (anti)kaon condensate), the references, the maximum mass of the cold star, the radius of the maximum-mass star, and the radius of the $1.4\,M_\odot$ star. The last column indicates whether the instability along the constant-$M_b$ path occurs, which is discussed in Sec.~\ref{sec:PNSevolution}. The columns $\Delta M_{b,\mathrm{max}}^{\mathrm{interm}}$ and $\Delta M_{b,\mathrm{max}}^{\mathrm{cold}}$ give the difference between the maximum baryonic mass at the intermediate ($S/A=2$, $Y_L=0.1$) and at the cold stage, respectively, and that at the early stage ($S/A=1$, $Y_L=0.4$), a negative value indicating that the early maximum-mass star cannot be supported at that stage.}
\label{table:DelayedCollapse}
\begin{ruledtabular}
\begin{tabular}{llllcccccc}
EoS & Framework & Content & Refs. & $M_{\mathrm{max}}^{\mathrm{cold}}$ & $R(M_{\mathrm{max}}^{\mathrm{cold}})$ & $R(1.4\,M_\odot)$ & $\Delta M_{b,\mathrm{max}}^{\mathrm{interm}}$ & $\Delta M_{b,\mathrm{max}}^{\mathrm{cold}}$ & Const-$M_b$ path \\
    &           &         &       & $[M_\odot]$ & [km] & [km] & $[M_\odot]$ & $[M_\odot]$ & instability \\
\hline
\multicolumn{10}{c}{Purely nucleonic models} \\
\hline
BCPN                  & NL-RMF & $N$ & \cite{BarmanInvestigatingRoleNuclearParametersOscillationModesHotNS2025, HornickRelativisticParametrizationNeutronMatterNS2018} & 2.54 & 12.19 & 13.37 & $+0.112$ & $+0.205$ & No  \\
IUF                   & NL-RMF & $N$ & \cite{FattoyevRelativisticEffectiveInteractionNucleiGiantResonancesNS2010, HempelStatisticalModelEoS2010} & 1.95 & 11.34 & 12.71 & $+0.171$ & $+0.124$ & No  \\
QMC-RMF1              & NL-RMF & $N$ & \cite{AlfordRelativisticMFTheorisNSPhysicsBasedChiralEFT2022, AlfordTabulatedEoSInformedChiralEFT2023} & 1.95 & 10.23 & 11.79 & $+0.139$ & $+0.157$ & No  \\
QMC-RMF2              & NL-RMF & $N$ & \cite{AlfordRelativisticMFTheorisNSPhysicsBasedChiralEFT2022, AlfordTabulatedEoSInformedChiralEFT2023} & 2.04 & 10.46 & 11.94 & $+0.144$ & $+0.176$ & No  \\
QMC-RMF3              & NL-RMF & $N$ & \cite{AlfordRelativisticMFTheorisNSPhysicsBasedChiralEFT2022, AlfordTabulatedEoSInformedChiralEFT2023} & 2.15 & 10.69 & 12.16 & $+0.129$ & $+0.185$ & No  \\
QMC-RMF4              & NL-RMF & $N$ & \cite{AlfordRelativisticMFTheorisNSPhysicsBasedChiralEFT2022, AlfordTabulatedEoSInformedChiralEFT2023} & 2.22 & 11.02 & 12.27 & $+0.151$ & $+0.209$ & No  \\
SFHo                  & NL-RMF & $N$ & \cite{SteinerCoreCollapseSupernovaEoS2013, HempelStatisticalModelEoS2010} & 2.06 & 10.32 & 11.91 & $+0.128$ & $+0.179$ & No  \\
SFHx                  & NL-RMF & $N$ & \cite{SteinerCoreCollapseSupernovaEoS2013, HempelStatisticalModelEoS2010} & 2.13 & 10.78 & 12.01 & $+0.151$ & $+0.200$ & No  \\
STOS(TM1)             & NL-RMF & $N$ & \cite{ShenRelativisticEoSSupernova1998, ShenNewEoSAstrophysicalSimulations2011, ShenRelativisticEOSCoreCollapseSupernovaSimulations2011} & 2.22 & 12.59 & 14.54 & $+0.319$ & $+0.200$ & No  \\
TM1e                  & NL-RMF & $N$ & \cite{SumiyoshiInfluenceDensityDependenceSymmetryEnergyHotDenseMatterSupernovaSimulations2019} & 2.14 & 12.01 & 13.44 & $+0.178$ & $+0.149$ & No  \\
DD2                   & DD-RMF & $N$ & \cite{TypelCompositionThermodynamicsNuclearMatterLightClusters2010, HempelStatisticalModelEoS2010} & 2.42 & 11.90 & 13.27 & $+0.080$ & $+0.169$ & No  \\
DDLS($L$=30)N         & DD-RMF & $N$ & \cite{TsiopelasFiniteTEoSCompactStarsY2024} & 2.49 & 12.04 & 12.89 & $+0.125$ & $+0.230$ & No  \\
DDLS($L$=50)N         & DD-RMF & $N$ & \cite{TsiopelasFiniteTEoSCompactStarsY2024} & 2.48 & 12.05 & 13.21 & $+0.106$ & $+0.206$ & No  \\
DDLS($L$=70)N         & DD-RMF & $N$ & \cite{TsiopelasFiniteTEoSCompactStarsY2024} & 2.46 & 12.12 & 13.59 & $+0.089$ & $+0.178$ & No  \\
RB2                   & Skyrme & $N$ & \cite{BeznogovBayesiaInferenceDenseMatterEoSExtendedSkyrmeInteractions2024, RadutaNewAbInitioSkyrmeEoSSimulationsNSSubsaturationDensityDomain2025, RadutaNewAbInitioSkyrmeEoSSimulationsNSSuprasaturationDensityDomain2026} & 2.17 & 10.87 & 12.57 & $+0.038$ & $+0.120$ & No  \\
RB5                   & Skyrme & $N$ & \cite{BeznogovBayesiaInferenceDenseMatterEoSExtendedSkyrmeInteractions2024, RadutaNewAbInitioSkyrmeEoSSimulationsNSSubsaturationDensityDomain2025, RadutaNewAbInitioSkyrmeEoSSimulationsNSSuprasaturationDensityDomain2026} & 2.03 & 10.04 & 12.09 & $+0.095$ & $+0.021$ & No  \\
RB6                   & Skyrme & $N$ & \cite{BeznogovBayesiaInferenceDenseMatterEoSExtendedSkyrmeInteractions2024, RadutaNewAbInitioSkyrmeEoSSimulationsNSSubsaturationDensityDomain2025, RadutaNewAbInitioSkyrmeEoSSimulationsNSSuprasaturationDensityDomain2026} & 2.05 & 10.17 & 12.03 & $+0.080$ & $+0.172$ & No  \\
RB7                   & Skyrme & $N$ & \cite{BeznogovBayesiaInferenceDenseMatterEoSExtendedSkyrmeInteractions2024, RadutaNewAbInitioSkyrmeEoSSimulationsNSSubsaturationDensityDomain2025, RadutaNewAbInitioSkyrmeEoSSimulationsNSSuprasaturationDensityDomain2026} & 2.05 & 10.26 & 12.25 & $+0.055$ & $+0.115$ & No  \\
RB8                   & Skyrme & $N$ & \cite{BeznogovBayesiaInferenceDenseMatterEoSExtendedSkyrmeInteractions2024, RadutaNewAbInitioSkyrmeEoSSimulationsNSSubsaturationDensityDomain2025, RadutaNewAbInitioSkyrmeEoSSimulationsNSSuprasaturationDensityDomain2026} & 2.08 & 10.34 & 12.26 & $+0.053$ & $+0.155$ & No  \\
SRO(SLy4)             & Skyrme & $N$ & \cite{SchneiderOpenSourceNuclearEoS2017, ChabanatSkyrmeParametrizationSubnnuclearNSDensities1998} & 2.06 & 9.99 & 11.73 & $+0.188$ & $+0.171$ & No  \\
FTNS                  & Variational & $N$ & \cite{TogashiNuclearEoSCoreCollapseSN2017, FurusawaNewEoSCCSN2017} & 2.23 & 10.24 & 11.70 & $+0.177$ & $+0.257$ & No  \\
SRO(APR)              & Variational & $N$ & \cite{SchneiderAPREoSSimulationsSupernovae2019, AkmalEosNucleonMatterNeutronStar1998} & 2.20 &  10.02 & 11.57 & $+0.119$ & $+0.221$ & No  \\
LPB                   & BHF & $N$ & \cite{LogotetaMicroscopicEoSHotNuclearMatter2021} & 2.10 & 10.40 & 12.40 & $+0.239$ & $+0.213$ & No  \\
\hline
\multicolumn{10}{c}{Models with exotic degrees of freedom} \\
\hline
BCPY                  & NL-RMF & $NY$ & \cite{BarmanInvestigatingRoleNuclearParametersOscillationModesHotNS2025, HornickRelativisticParametrizationNeutronMatterNS2018} & 2.13 & 11.68 & 13.37 & $-0.128$ & $-0.064$ & Yes \\
FSU2H$^{*}$           & NL-RMF & $NY$ & \cite{KochankovskiEoSHotHyperonicNSM2022, KochankovskiHyperonicUncertaintiesNS2024} & 2.01 & 11.98 & 13.19 & $-0.137$ & $-0.084$ & Yes \\
SFHoY                 & NL-RMF & $NY$ & \cite{FortinHyperonsHotDenseMatter2018} & 1.99 & 10.32 & 11.91 & $+0.025$ & $+0.105$ & No  \\
STOS(TM1)Y            & NL-RMF & $N\Lambda$ & \cite{ShenRelativisticEoSSupernova1998, ShenNewEoSAstrophysicalSimulations2011, ShenRelativisticEOSCoreCollapseSupernovaSimulations2011} & 1.76 & 13.13 & 14.55 & $-0.029$ & $-0.131$ & Yes \\
BHB$\Lambda\phi$      & DD-RMF & $N\Lambda$ & \cite{BanikProbingMetastabiliyPNS2014} & 2.10 & 11.61 & 13.24 & $-0.032$ & $+0.016$ & Yes \\
BHB$\Lambda\phi$K     & DD-RMF & $N\Lambda K$ & \cite{MalikEoSHyperonsKaonsSupernovaNSMergers2021} & 1.94 & 11.89 & 13.24 & $-0.138$ & $-0.202$ & Yes \\
DD2K                  & DD-RMF & $NK$ & \cite{MalikNewEoSBECondensateAntikaonSupernovaNSMergerSimulations2021} & 2.21 & 12.06 & 13.24 & $-0.050$ & $-0.076$ & Yes \\
DD2Y                  & DD-RMF & $NY$ & \cite{MarquesTemperatureDependentHEoS2014} & 2.04 & 11.44 & 13.25 & $-0.132$ & $-0.066$ & Yes \\
DD2Y$\Delta$(1.1,1.1) & DD-RMF & $NY\Delta$ & \cite{RadutaEoSHotNSExoticParticleDoF2022} & 2.04 & 11.29 & 13.00 & $-0.122$ & $-0.051$ & Yes \\
DD2Y$\Delta$(1.2,1.1) & DD-RMF & $NY\Delta$ & \cite{RadutaEoSHotNSExoticParticleDoF2022} & 2.06 & 11.10 & 12.31 & $-0.087$ & $+0.010$ & Yes \\
DD2Y$\Delta$(1.2,1.3) & DD-RMF & $NY\Delta$ & \cite{RadutaEoSHotNSExoticParticleDoF2022} & 2.04 & 11.44 & 13.25 & $-0.127$ & $-0.071$ & Yes \\
DDLS($L$=30)Y         & DD-RMF & $NY$ & \cite{TsiopelasFiniteTEoSCompactStarsY2024} & 2.01 & 11.74 & 12.89 & $-0.173$ & $-0.122$ & Yes \\
DDLS($L$=50)Y         & DD-RMF & $NY$ & \cite{TsiopelasFiniteTEoSCompactStarsY2024} & 1.99 & 11.75 & 13.21 & $-0.190$ & $-0.149$ & Yes \\
DDLS($L$=70)Y         & DD-RMF & $NY$ & \cite{TsiopelasFiniteTEoSCompactStarsY2024} & 1.97 & 11.75 & 13.59 & $-0.208$ & $-0.181$ & Yes \\
DNS                   & CMF & $NY$ & \cite{DexheimerTabulatedNSEoSCMF2017} & 2.09 & 12.25 & 14.28 & $+0.102$ & $+0.177$ & No  \\
QMC-A                 & QMC & $NY$ & \cite{StoneEoSHotYMatterQMCA2021} & 1.99 & 12.07 & 12.87 & $-0.101$ & $+0.013$ & Yes \\
\end{tabular}
\end{ruledtabular}
\end{table*}

\section{PNS evolution of maximum-mass stars}
\label{sec:PNSevolution}

We first test, model-by-model, the delayed collapse scenario put forward three decades ago~\cite{BrownBetheScenarioLowMassBHGalaxy1994, ThorssonCompositionStructureEvolutionNSKaonCondensates1994, KeilJankaHadronicPhaseTransitionsDelayedCollapse1995, GlendenningPromptSubsidencePNSBlackHole1995, BombaciMmaxNS1996}, namely that, in the absence of accretion, the most massive exotic-rich stars are bound to collapse during their PNS evolution at constant baryonic mass $M_b$. The mechanism, examined in detail in \cref{sec:deleptonization}, is the following. In the early, lepton-rich stage, the appearance of hyperons is suppressed, so that the EoS stays comparatively stiff and the maximum mass is large. The subsequent deleptonization and heating of the core allow the hyperons to populate the medium, which softens the EoS and lowers the maximum mass that can be sustained. As a consequence, a star born close to the maximum-mass configuration may no longer find a stable configuration of the same baryonic mass at later times, as the maximum-mass configuration at that stage might have a smaller baryonic mass,  thus leading to a delayed collapse into a black hole.

In principle, a proper analysis should follow a full PNS evolution for each EoS, coupling the stellar structure equations to the neutrino transport and closing the system with a finite-temperature, lepton-dependent EoS~\cite{Pascal:2022qeg, Hudepohl:2009tyy, Fischer:2009af}. Such simulations show that the lepton fraction and the temperature profiles vary, at any given instant, along the radial coordinate of the star. However, since these simulations are computationally extremely expensive, we adopt the commonly used simplified approach (see, for example, Refs.~\cite{PrakashCompositionStructurePNS1997, BurgioHyperonStarsFiniteTBruecknerTheory2011, RadutaEoSHotNSExoticParticleDoF2022}) of considering a constant total lepton fraction $Y_L = Y_e + Y_\nu$ and a constant total entropy per baryon $S/A$ (in units of $k_B$, including the neutrino contribution) throughout the star\footnote{The EoSs employed in this work, all taken from the \textsc{CompOSE} database~\cite{TypelCompOSE2015, TypelCompOSEManual2022}, do not include the contribution of neutrinos, which we therefore add explicitly. Electron neutrinos and antineutrinos are treated as a massless free Fermi gas, whose chemical potential follows from the weak equilibrium condition $\mu_n + \mu_{\nu_e} = \mu_p + \mu_e$, and their contributions to the pressure, energy density and entropy are added to the tabulated matter quantities. Below $\rho_B = 10^{-3}\,\mathrm{fm}^{-3}$, where the neutrinos are no longer trapped, their contribution is dropped by setting the neutrino density to zero, $\rho_\nu = 0$, and the charged-lepton fraction is kept at the value of $Y_L$. The DD2 and QMC-A tables contain muons in addition to electrons, with $\mu_\mu = \mu_e$, and for these two models the lepton fraction reads $Y_L = Y_e + Y_\mu + Y_\nu$. The contribution of the muon neutrinos is neglected for simplicity.}. This approximation is reasonable at a few representative instants of the star's evolution, such as the following three characteristic snapshots~\cite{PonsEvolutionPNS1999}:

\begin{enumerate}
    \item the earliest PNS stage, in which the matter retains a high lepton
    fraction at a moderate entropy per baryon, $Y_L = 0.4$ and $S/A = 1$;
    \item an intermediate, nearly deleptonized stage,
    $Y_L = 0.1$ and $S/A = 2$;
    \item the fully evolved cold neutron star, in $\beta$-equilibrium at
    $T = 0$.
\end{enumerate}

The neutrinos remain trapped in the hot core at both finite-temperature stages, so that the first two snapshots differ in the net lepton number that has already diffused out of the star and in their entropy per baryon. In Refs.~\cite{PrakashCompositionStructurePNS1997,PonsEvolutionPNS1999} the end of the deleptonization is characterized by a vanishing net neutrino number, $\mu_{\nu_e}=0$, which corresponds to a lepton fraction that depends on the density and on the model. Here we adopt the constant value $Y_L=0.1$ as a proxy for this stage instead, since it allows us to describe the whole evolution with the same pair of variables, $Y_L$ and $S/A$, that are varied separately in \cref{sec:cooling_deleptonization}.  Note that the fully deleptonized stage of the proto-neutron star evolution, i.e.\ where all neutrinos have     left the star, still has a nonvanishing lepton fraction due to the presence of electrons, so $Y_L=0$ cannot be realized even for the final cold neutron star.

The baryonic mass of a given configuration is obtained from the conservation of baryon number together with the solution of the Tolman-Oppenheimer-Volkoff equations as
\begin{equation}
\label{eq:baryonic_mass}
M_b = m_N \int_0^R \frac{4\pi r^2\, \rho_B(r)}{\sqrt{1 - 2Gm(r)/(rc^2)}}\, dr ,
\end{equation}
where $m(r)$ is the gravitational mass enclosed within the radius $r$ and $m_N = 939$~MeV is the nucleon mass. For each EoS, we compute the $M(R)$ curves at these three stages. We then locate the maximum-baryonic-mass configuration at $Y_L = 0.4$, $S/A = 1$, and check whether a stable stellar configuration with the same baryonic mass still exists at the two later stages by looking at the maximum baryonic mass in those cases. If no such configuration exists, the star is bound to collapse along the constant-$M_b$ trajectory, providing a clean realization of the delayed collapse scenario. The outcome for each EoS is given in the last column of \cref{table:DelayedCollapse}. \cref{table:DelayedCollapse} also lists the differences $\Delta M_{b,\mathrm{max}}$ between the maximum baryonic mass at each of the two later stages and that at the early stage, $\Delta M_{b,\mathrm{max}} = M_{b,\mathrm{max}} \text{(later stage)} - M_{b,\mathrm{max}}\text{(early stage)}$, a negative value indicating that the early maximum-mass star cannot be supported at that stage.

Out of the 16 EoSs that contain exotic hadrons, 14 predict an instability along the constant-$M_b$ path, i.e.~a delayed collapse of the most massive PNS into a black hole during its evolution. The only two exceptions are the SFHoY and DNS models.  For the collapsing models, the maximum baryonic mass decreases by $0.03$ to $0.21\,M_\odot$ with respect to the early stage. This mass difference defines the width of the window for which a PNS is stable at birth and collapses later. For most of the models with exotic hadrons, the largest difference is found at the hot intermediate stage, and not in the cold star, and for three of them, BHB$\Lambda\phi$, DD2Y$\Delta$(1.2,1.1) and QMC-A, the instability appears only at that stage, since their cold maximum baryonic mass is slightly larger than the early-stage value $M_{b,\mathrm{max}}$. By contrast, none of the purely nucleonic EoSs in our sample predicts such an instability, since every configuration with the same baryonic mass as the initial maximum-mass star survives at both later stages, and the star reaches a stable cold configuration. While the validity of the delayed-collapse scenario had already been tested in previous works, those studies were restricted to one or a few EoSs (see, e.g., Refs.~\cite{RadutaPNSHeavyBaryons2020,BanikProbingMetastabiliyPNS2014}), so that a systematic confirmation across a large and heterogeneous set of models was still lacking. Our analysis fills this gap and shows that the delayed collapse of the most massive exotic-rich PNS is a robust and rather generic outcome.

To characterize the global content of exotic species of these configurations, we use the mass-averaged fractions, defined for a generic species fraction $Y_i$ as
\begin{equation}
\label{eq:mass_average}
\langle Y_i \rangle = \frac{1}{M}\int Y_i \, dM ,
\end{equation}
where the integral runs over the whole star and weights the local fraction $Y_i$ by the amount of gravitational mass ($M$) it represents.

We now follow the mass-averaged hyperon fraction $\langle Y_{\mathrm{hyp+\Delta}} \rangle$ through the three stages, counting the $\Delta$ resonances as well for the models that include them. Since for most of the hyperonic EoSs a star at the early maximum baryonic mass $M_{b,\mathrm{max}}$ no longer exists at the later stages, we evaluate the fractions for the maximum-mass star that is physically realizable at each given snapshot. In the same way, we also report the mass-averaged kaonic fraction at the different stages of the evolution. The results are collected in \cref{table:hyp_kao_fractions}. 

The models that include $\Delta$ resonances systematically display some of the highest mass-averaged hyperon fractions, reaching $\langle Y_{\mathrm{hyp+\Delta}} \rangle > 0.35$ in the intermediate phase for the maximum-mass star obtained with DD2Y$\Delta$(1.2,1.1). The two hyperonic EoSs that do not predict a delayed collapse, SFHoY and DNS, are found to have, together with QMC-A, the lowest mass-averaged hyperon fractions of the whole set.
 For example, the DNS model predicts almost no hyperons during the early phase ($\langle Y_{\mathrm{hyp+\Delta}} \rangle\approx0.013$) and only a small fraction in the cold evolved star ($\langle Y_{\mathrm{hyp+\Delta}} \rangle\approx0.045$).
The hyperon fraction by itself, however, does not determine whether the delayed collapse takes place, as the QMC-A model illustrates. In this model, the $\Lambda$ and $\Xi^-$ hyperons appear at almost the same, comparatively high density, $\rho_B \simeq 0.5$--$0.6\,\mathrm{fm}^{-3}$~\cite{StoneEoSHotYMatterQMCA2021}, so that they are confined to the innermost core and their mass-averaged fraction remains small. The instability appears only at the hot intermediate stage, where $\langle Y_{\mathrm{hyp+\Delta}} \rangle$ is more than four times larger than in the cold star, and it can be traced back to the thermal behavior of the model, whose thermal index stays low over a wide range of densities once the hyperons are thermally produced~\cite{KochankovskiImpactHyperonsNSMergers2025}. The heated core therefore gains little thermal pressure while it loses degeneracy pressure, and the maximum baryonic mass falls below that of the early stage, whereas the cold star, with its small hyperon content, is again able to support it.

We now turn to the EoSs that include a kaon condensate, where even a small kaonic fraction compared with the typical hyperonic ones can trigger a delayed collapse. Indeed, the DD2K model, which collapses during its evolution and includes kaons as the only exotic degree of freedom, predicts a mass-averaged kaonic fraction that grows along the whole evolution and reaches its maximum in the cold star, $\langle Y_K \rangle\approx0.1$, whereas for most of the hyperonic EoSs the maximum hyperon fraction is attained in the intermediate phase. The growth of the kaon content is governed by the condition $\mu_{K^-}=\mu_n-\mu_p=\mu_e-\mu_{\nu_e}$, i.e.~by the isospin asymmetry of the medium. In the early, lepton-rich stage, the matter is relatively isospin-symmetric and $\mu_n-\mu_p$ is small, so that the appearance of $K^-$ is disfavored. As the star deleptonizes, the matter becomes increasingly neutron-rich, $\mu_n-\mu_p$ grows and the kaon population increases, as already pointed out in Ref.~\cite{PonsEvolutionPNSKaonCondensates2001}. The same mechanism also favors the hyperons, but the different behavior of the two species arises from their response to the temperature. The hyperons are thermally populated well below their cold threshold, so that their abundance is the largest in the hot intermediate stage and decreases as the star cools. The condensate, by contrast, occupies the zero-momentum state and is therefore suppressed rather than enhanced by the temperature, so that the kaon content keeps growing as the remaining net neutrinos leave the star and the matter cools towards the catalyzed configuration. A particularly clean illustration is provided by the BHB$\Lambda\phi$ and BHB$\Lambda\phi$K models, which share the same hyperonic base and differ only by the inclusion of the kaon condensate. As can be seen in \cref{table:hyp_kao_fractions}, for BHB$\Lambda\phi$K the hyperon fraction peaks in the intermediate phase, while its kaonic fraction grows monotonically and reaches its maximum in the cold star, exactly as for DD2K.

Finally, we comment on the limitations of the present approach. First, by following the evolution along sequences of fixed lepton fraction and entropy per baryon, we neglect the radial dependence of these quantities, as well as the effect of rotation. The latter has recently been shown to modify the maximum mass appreciably only for stars rotating close to the Kepler frequency, while at lower frequencies its impact on the evolution of $M_{\mathrm{max}}$ remains modest~\cite{daSilvaHyperonsDeltaResonancesRotatingPNS2025}. Second, our stability criterion assumes that the composition remains in weak equilibrium at every instant. If, instead, some of the relevant reactions are slow compared with the dynamical timescale~\cite{ZatiniLambdaHyperonsCCSNEquilibrium2026}, the composition can be regarded as frozen, in which case stable configurations may persist somewhat beyond the catalyzed maximum mass, an effect that becomes more pronounced for the compositionally richer models with several hyperonic and $\Delta$ species~\cite{CanullanPascualNSStabilityBeyondMassPeakOutEquilibriumPerturbations2025}. These refinements would quantitatively shift the boundaries of the metastability window, but they are not expected to alter the qualitative conclusion on whether a given EoS leads to a delayed collapse. We also recall that all the hyperonic EoSs considered here are phenomenological, since no microscopic finite-temperature tables with hyperons are currently available, so that the generality of our conclusion is to be understood within this class of models.

\begin{table}[t]
\caption{Mass-averaged hyperon (plus $\Delta$) fraction $\langle Y_{\mathrm{hyp+\Delta}} \rangle$ and kaon
fraction $\langle Y_K \rangle$ of the maximum-mass star for the
hyperonic and kaonic EoSs, at the three snapshots of the
proto-neutron-star evolution: Early ($S/A=1$, $Y_L=0.4$),
Intermediate ($S/A=2$, $Y_L=0.1$) and Cold ($\beta$-equilibrium).
A slash ($/$) denotes a species absent from the model.}
\label{table:hyp_kao_fractions}
\begin{ruledtabular}
\begin{tabular}{lcccccc}
 & \multicolumn{3}{c}{$\langle Y_{\mathrm{hyp+\Delta}} \rangle$} & \multicolumn{3}{c}{$\langle Y_K \rangle$} \\
\cline{2-4} \cline{5-7}
EoS & Early & Interm. & Cold & Early & Interm. & Cold \\
\hline
BCPY                     & 0.124 & 0.220 & 0.205 & $/$ & $/$ & $/$ \\
BHB$\Lambda\phi$         & 0.120 & 0.212 & 0.203 & $/$ & $/$ & $/$ \\
DD2Y                     & 0.113 & 0.218 & 0.203 & $/$ & $/$ & $/$ \\
DD2Y$\Delta$(1.1,1.1)    & 0.114 & 0.269 & 0.271 & $/$ & $/$ & $/$ \\
DD2Y$\Delta$(1.2,1.1)    & 0.168 & 0.357 & 0.350 & $/$ & $/$ & $/$ \\
DD2Y$\Delta$(1.2,1.3)    & 0.108 & 0.221 & 0.198 & $/$ & $/$ & $/$ \\
DDLS($L$=30)Y            & 0.097 & 0.194 & 0.172 & $/$ & $/$ & $/$ \\
DDLS($L$=50)Y            & 0.098 & 0.202 & 0.185 & $/$ & $/$ & $/$ \\
DDLS($L$=70)Y            & 0.101 & 0.213 & 0.202 & $/$ & $/$ & $/$ \\
SFHoY                    & 0.015 & 0.082 & 0.039 & $/$ & $/$ & $/$ \\
STOS(TM1)Y               & 0.081 & 0.151 & 0.120 & $/$ & $/$ & $/$ \\
DNS                      & 0.013 & 0.113 & 0.045 & $/$ & $/$ & $/$ \\
FSU2H$^{*}$              & 0.087 & 0.180 & 0.146 & $/$ & $/$ & $/$ \\
QMC-A                    & 0.014 & 0.090 & 0.021 & $/$ & $/$ & $/$ \\
\hline
BHB$\Lambda\phi$K        & 0.106 & 0.174 & 0.145 & 0.004 & 0.049 & 0.061 \\
DD2K                     & $/$ & $/$ & $/$ & 0.015 & 0.075 & 0.098 \\
\end{tabular}
\end{ruledtabular}
\end{table}
\section{Deleptonization and cooling of exotic-rich and nuclear matter}
\label{sec:cooling_deleptonization}

The analysis presented in Sec.~\ref{sec:PNSevolution} has confirmed that the EoSs containing exotic hadrons generically drive the most massive PNSs towards a delayed collapse along a constant baryonic mass trajectory, whereas none of the purely nucleonic models do. During the Kelvin-Helmholtz epoch, however, two processes act on matter at the same time, namely its deleptonization and the change of its thermal content, with the heating of the newly born star followed by its cooling. Since both modify the stiffness of the EoS, the results of the previous section alone cannot tell which of them drives the instability. In the present section we therefore study the two processes separately. We first fix the entropy per baryon and vary the total lepton fraction, and then fix the lepton fraction and vary the entropy, so that the effect of each process on the maximum-mass configurations can be assessed on its own.

\subsection{Deleptonization}
\label{sec:deleptonization}

To isolate the effect of deleptonization, we fix the entropy per baryon to $S/A = 1$ and let the total lepton fraction $Y_L$
  decrease from the value that characterizes the early, lepton-rich stage down to the lepton fraction of the intermediate stage of the evolution\footnote{We present the results obtained at $S/A = 1$. We have verified that the qualitative behavior is unchanged at different entropies.}. The thermal content of the matter is then fixed, and the changes in the maximum-mass configuration follow mainly from the loss of leptons. Note that this is an academic exercise to disentangle the effect of the lepton fraction alone, since in a real PNS the escaping neutrinos also carry entropy away. Moreover, a fixed entropy per baryon does not correspond to a fixed temperature. Since the total entropy per baryon includes the neutrino contribution, the entropy that is no longer carried by the trapped neutrinos is transferred to the matter along the sequence, and the temperature of the matter therefore rises. This rise is considerably smaller for the hyperonic models, as discussed in Sec.~\ref{sec:other_results}.

\begin{figure}[tbp]
    \centering
    \includegraphics[width=0.99\columnwidth]{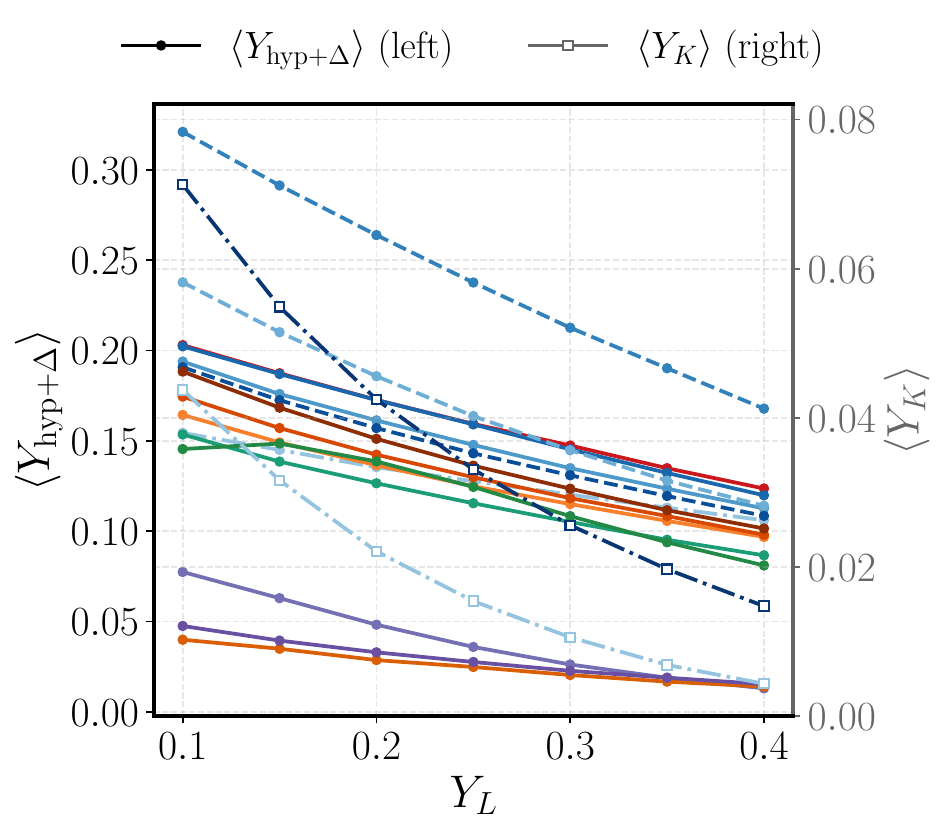}
    \caption{Mass-averaged hyperon (plus $\Delta$) fraction $\langle Y_{\mathrm{hyp+\Delta}} \rangle$ (left axis, filled
circles) and kaon fraction $\langle Y_K \rangle$ (right axis, open squares)
of the maximum-mass star as a function of the total lepton fraction $Y_L$, for the EoSs with
exotic degrees of freedom at fixed entropy per baryon $S/A = 1$. The kaon fraction
$\langle Y_K \rangle$ is shown for the two kaonic EoSs, DD2K and BHB$\Lambda\phi$K.
The line styles and colors are the same as in \cref{fig:cold_MR}.}
    \label{fig:hyp_frac_YL}
\end{figure}

Since the response of the hyperonic EoSs is tied to the hyperon content of the medium, we begin with \cref{fig:hyp_frac_YL}, which shows the mass-averaged hyperon fraction $\langle Y_{\mathrm{hyp+\Delta}}\rangle$ and kaon fraction $\langle Y_{\mathrm{K}} \rangle$ of the maximum-mass star, defined in \cref{eq:mass_average}, as a function of the lepton fraction. Here $\langle Y_{\mathrm{hyp+\Delta}}\rangle$ is the mass average of the summed $\Lambda$, $\Sigma$ and $\Xi$ fractions, to which the $\Delta$ fraction is added for the models that include $\Delta$ resonances. The values at $Y_L = 0.4$ coincide with the early-stage entries of \cref{table:hyp_kao_fractions}. For all the models except STOS(TM1)Y, whose hyperon fraction saturates below $Y_L \approx 0.15$, $\langle Y_{\mathrm{hyp+\Delta}} \rangle$ increases monotonically as the lepton fraction decreases. For most of the models, it grows by a factor of about 1.5 to 2 between $Y_L = 0.4$ and $Y_L = 0.1$, while for SFHoY, DNS and QMC-A, which have the smallest hyperon content, the relative increase is even larger, by a factor of 3 to 6. This is the expected behavior, since a high lepton fraction keeps the chemical potentials of the baryons relatively low and, at the same time, forces matter to be more isospin-symmetric, both effects hindering the onset and the population of the hyperonic degrees of freedom~\cite{PrakashCompositionStructurePNS1997, VidanaMicroscopicStudyNeutrinoTrappingHyperonStars2003, BurgioHyperonStarsFiniteTBruecknerTheory2011}. The isospin argument is particularly relevant for the negatively charged species, such as the $\Sigma^-$, the $\Xi^-$, and the $\Delta^-$, which are favored by the neutron excess, and it is consistent with the fact that two of the models including $\Delta$ resonances, DD2Y$\Delta$(1.1,1.1) and DD2Y$\Delta$(1.2,1.1), display the largest increase of $\langle Y_{\mathrm{hyp+\Delta}} \rangle$ during the deleptonization. As the trapped neutrinos gradually leave the star and $Y_L$ decreases, this suppression is progressively lifted, and the hyperons are allowed to populate the core, so that the hyperon content of the maximum-mass star is systematically larger in the more deleptonized configurations. The same qualitative pattern describes the trend of the kaonic fraction.

\begin{figure*}[tp]
    \centering
    \subfigure[Hyperonic EoSs]{\includegraphics[width=0.45\textwidth]{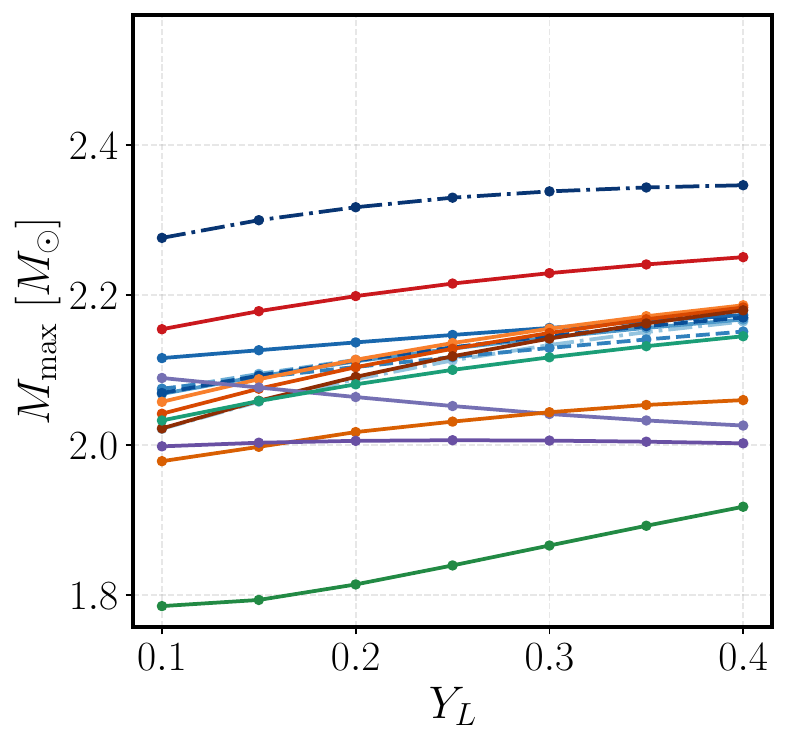}}
    \hfill
    \subfigure[Nucleonic EoSs]{\includegraphics[width=0.45\textwidth]{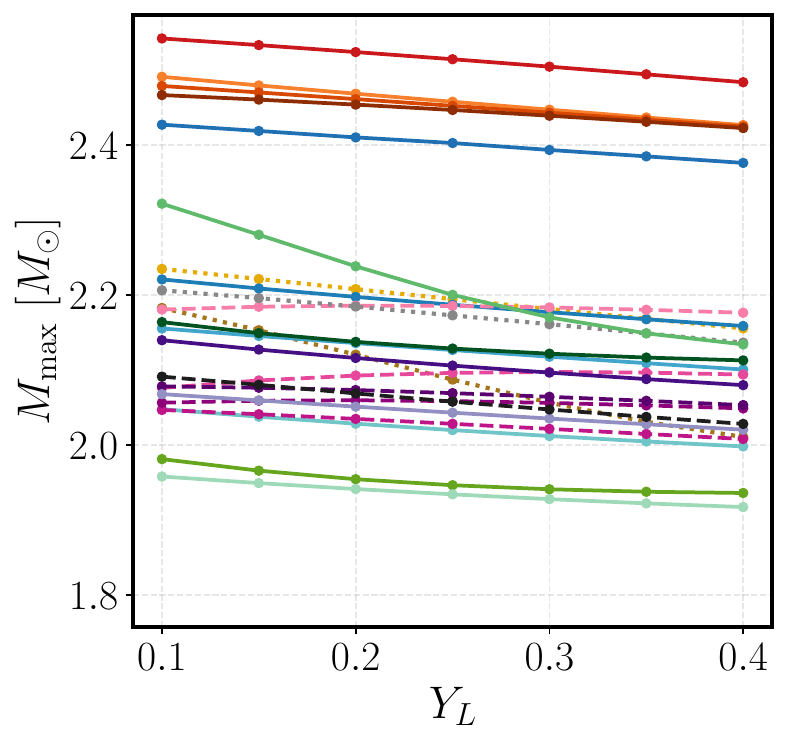}}
    \caption{Maximum gravitational mass $M_{\mathrm{max}}$ as a function of the
    total lepton fraction $Y_L$, at fixed entropy per baryon $S/A = 1$, for the
    hyperonic EoSs (left panel) and the nucleonic EoSs (right panel). The line
    styles and colors are the same as in \cref{fig:cold_MR}.}
    \label{fig:grav_mass_YL}
\end{figure*}

The consequences of this behavior are most directly seen in the evolution of the maximum gravitational mass, which we display in \cref{fig:grav_mass_YL} for the hyperonic EoSs (left panel) and for the nucleonic EoSs (right panel). For most of the hyperonic models, and in particular for every hyperonic EoS that was found to predict a delayed collapse in \cref{sec:PNSevolution}, the maximum gravitational mass decreases as the lepton fraction is reduced. This is the well-known softening of the EoS induced by the appearance of hyperons~\cite{GlendenningHyperonCompositionNS1982, ChatterjeeHyperonsExistNS2016, SchaffnerBielichCompactStarPhysicsbook2020}. The deleptonization enriches the medium with additional hadronic species, which lower the pressure of the highly degenerate nucleons and leptons. The star is then less able to support itself against gravity, and its maximum mass falls.

For the nucleonic EoSs, the global trend is the opposite, and for most of the nucleonic models the maximum gravitational mass increases as the lepton fraction decreases, as already found in earlier studies of PNSs~\cite{PrakashCompositionStructurePNS1997}. The trapped leptons keep the matter relatively close to isospin symmetry, so that the contribution of the symmetry energy to the pressure remains small. As the neutrinos leave the star, the matter becomes more neutron-rich, and the pressure of the nucleons grows, more than compensating the loss of the lepton pressure, so that the EoS becomes stiffer and the maximum mass increases. The only exceptions are the Skyrme models RB2, RB5 and RB7, for which $M_{\mathrm{max}}$ remains almost constant or decreases slightly, although their maximum baryonic mass increases, as shown below. 

\begin{figure*}[tp]
    \centering
    \subfigure[Hyperonic EoSs]{\includegraphics[width=0.45\textwidth]{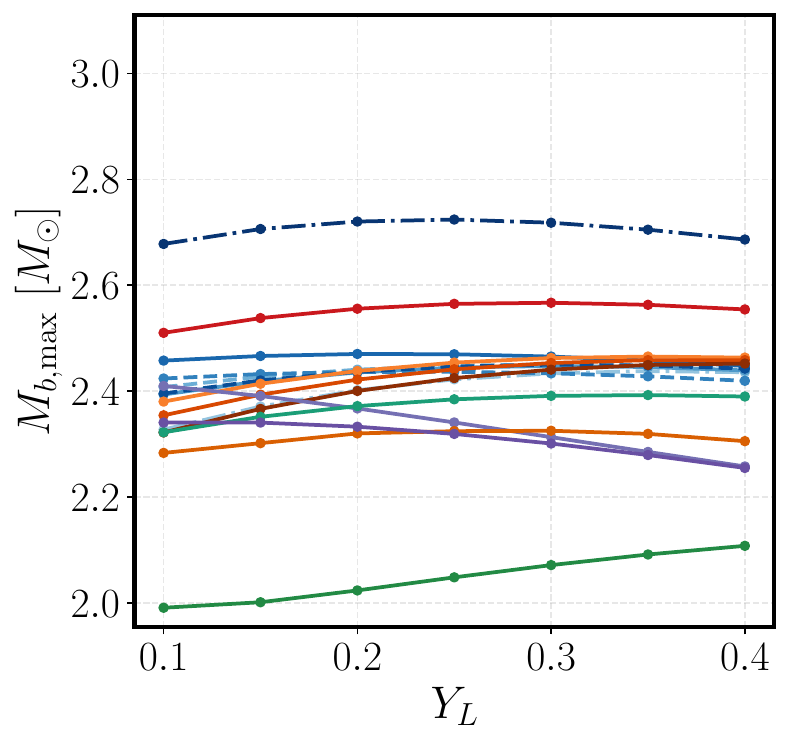}}
    \hfill
    \subfigure[Nucleonic EoSs]{\includegraphics[width=0.45\textwidth]{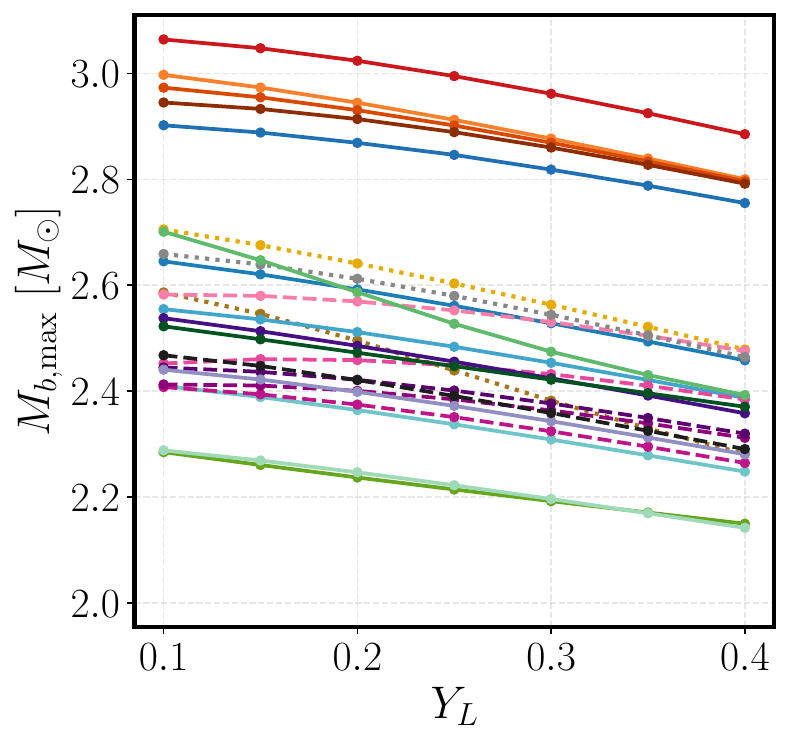}}
    \caption{Maximum baryonic mass $M_{b,\mathrm{max}}$ as a function of the
    total lepton fraction $Y_L$, at fixed entropy per baryon $S/A = 1$, for the
    hyperonic EoSs (left panel) and the nucleonic EoSs (right panel). The line
    styles and colors are the same as in \cref{fig:cold_MR}.}
    \label{fig:bar_mass_YL}
\end{figure*}

The behavior of the maximum gravitational mass is reflected in the maximum baryonic mass, shown in \cref{fig:bar_mass_YL}, and here the two families of models respond to the deleptonization in opposite directions. For every nucleonic EoS, the maximum baryonic mass increases as the star deleptonizes, by $0.07$ to $0.31\,M_\odot$ between $Y_L = 0.4$ and $Y_L = 0.1$. In the absence of exotic hadrons, the loss of leptons therefore always stiffens the EoS, since the growing isospin asymmetry raises the pressure of the nucleonic matter, as discussed above. When hyperons are present, this stiffening is suppressed. The appearance of hyperons in the core, shown in \cref{fig:hyp_frac_YL}, counteracts the stiffening provided by the isospin asymmetry, and for most of the hyperonic models it overcomes it, so that $M_{b,\mathrm{max}}$ reaches its largest value at intermediate lepton fractions and is lower at $Y_L = 0.1$ than at $Y_L = 0.4$. The difference $\Delta M_{b,\mathrm{max}} = M_{b,\mathrm{max}}(Y_L = 0.1) - M_{b,\mathrm{max}}(Y_L = 0.4)$ reaches $-0.13\,M_\odot$, i.e.\ the maximum baryonic mass is lower at $Y_L = 0.1$ by up to $0.13\,M_\odot$. The effect is most transparent when a hyperonic model is compared with its nucleonic counterpart. For the DD2 interaction, the deleptonization raises $M_{b,\mathrm{max}}$ by $0.15\,M_\odot$ in the purely nucleonic case, whereas it lowers it by $0.05\,M_\odot$ when hyperons are included, and the same behavior is found for the BCP, DDLS and STOS(TM1) families. The only hyperonic models for which the deleptonization still increases the maximum baryonic mass are SFHoY and DNS, which carry a small hyperon content and are precisely the two hyperonic EoSs that do not predict a delayed collapse. As we will show in \cref{sec:heating_cooling}, the heating of matter lowers $M_{b,\mathrm{max}}$ for both families by a comparable amount. These two families are therefore distinguished by the direction of their response to the deleptonization rather than by their response to the heating~\cite{BombaciMmaxNS1996, PrakashCompositionStructurePNS1997, PonsEvolutionPNS1999}.

\begin{figure*}[tp]
    \centering
    \subfigure[Hyperonic EoSs]{\includegraphics[width=0.45\textwidth]{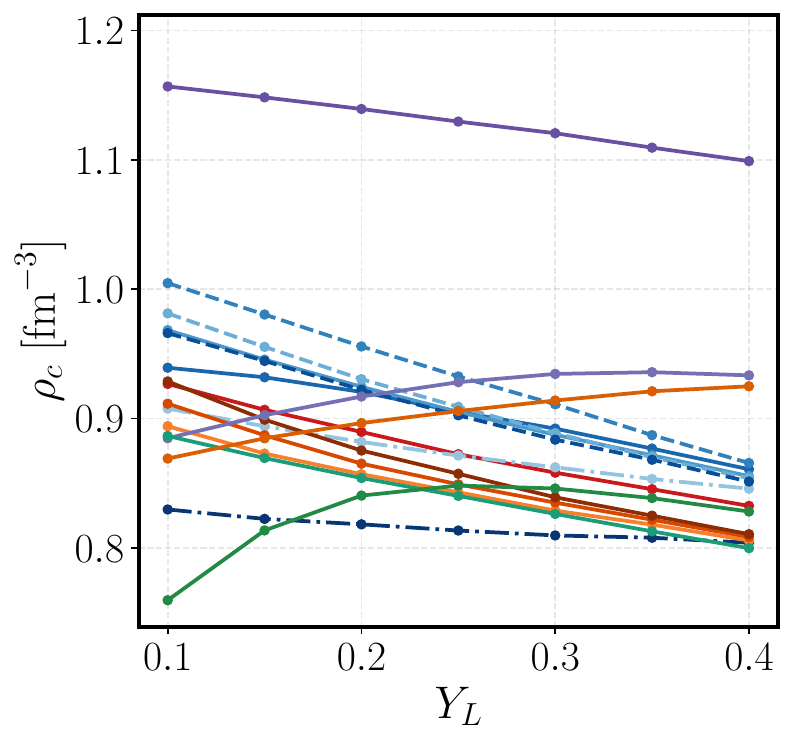}}
    \hfill
    \subfigure[Nucleonic EoSs]{\includegraphics[width=0.45\textwidth]{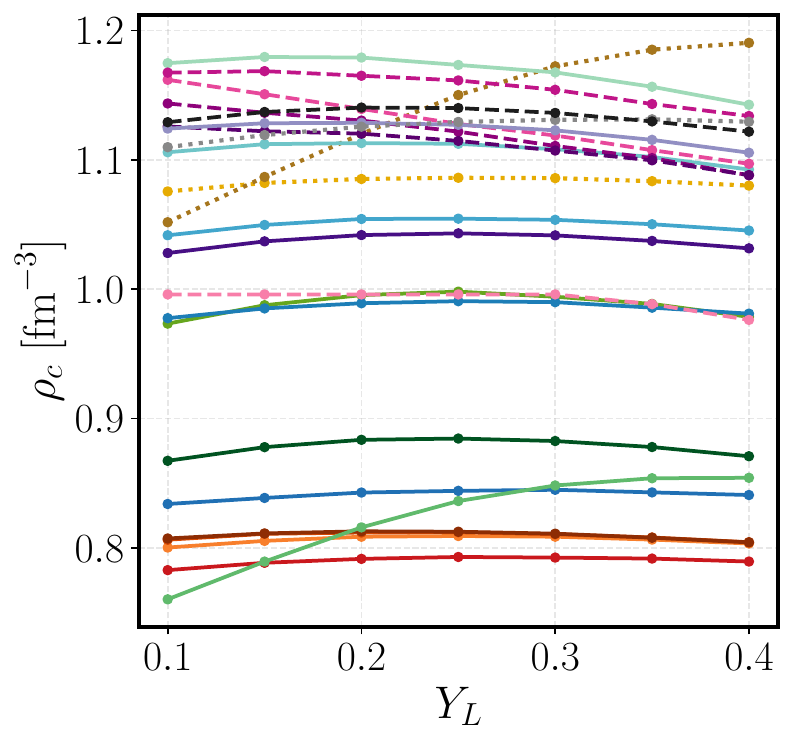}}
    \caption{Central density $\rho_c$ of the maximum-mass star as a function of
    the total lepton fraction $Y_L$, at fixed entropy per baryon $S/A = 1$, for
    the hyperonic EoSs (left panel) and the nucleonic EoSs (right panel). The
    line styles and colors are the same as in \cref{fig:cold_MR}.}
    \label{fig:central_density_YL}
\end{figure*}

We next consider the central density of the maximum-mass star as a function of the lepton fraction, shown in \cref{fig:central_density_YL}. Here too, the nucleonic and hyperonic families behave differently. For most of the hyperonic EoSs, the central density $\rho_c$ increases as the lepton fraction decreases, reflecting the fact that the softening of the EoS forces the maximum-mass configuration to become more compact. For most of the nucleonic EoSs, on the contrary, the central density of the maximum-mass star does not change significantly with the lepton fraction.

\begin{figure*}[tp]
    \centering
    \subfigure[Hyperonic EoSs]{\includegraphics[width=0.45\textwidth]{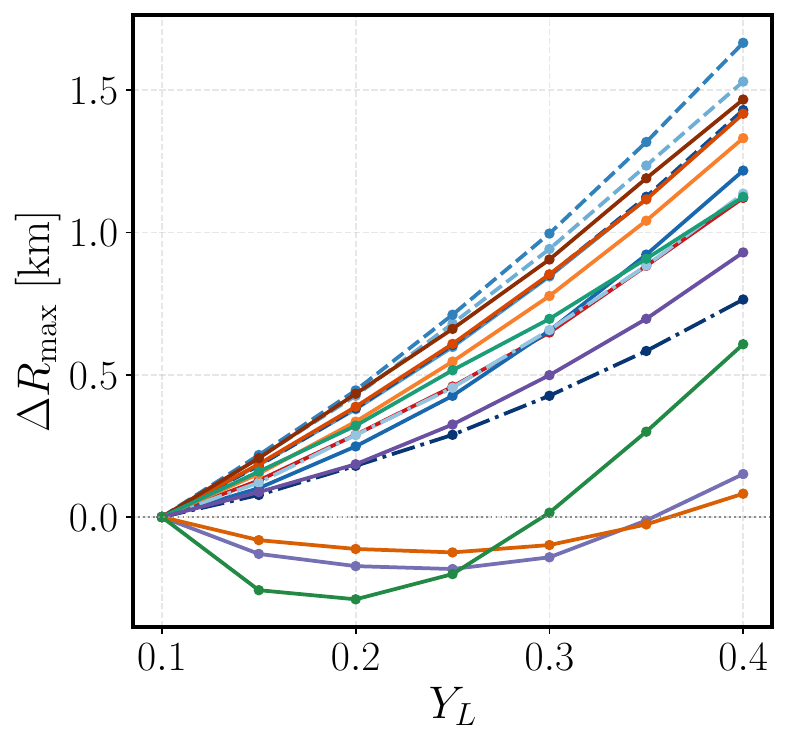}}
    \hfill
    \subfigure[Nucleonic EoSs]{\includegraphics[width=0.45\textwidth]{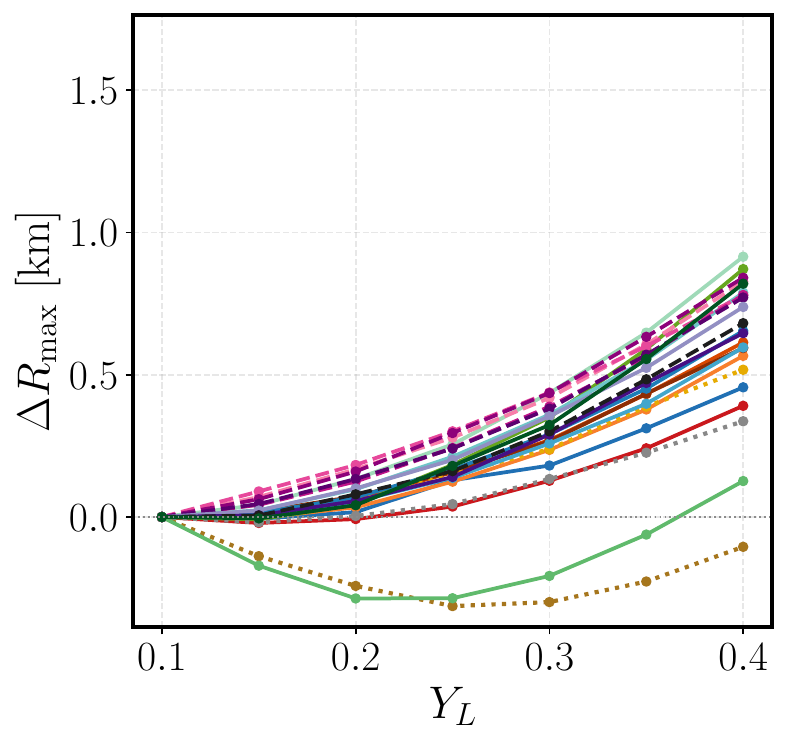}}
    \caption{Variation of the radius of the maximum-mass star,
    $\Delta R_{\max}$, defined in \cref{eq:delta_Rmax}, as a function of the total lepton fraction $Y_L$, at fixed entropy per baryon $S/A = 1$, for the hyperonic EoSs (left panel) and the nucleonic EoSs (right panel). The line styles and colors are the same as in \cref{fig:cold_MR}.}
    \label{fig:delta_Rmax_YL}
\end{figure*}

The increase in the central density found for the hyperonic models reflects a contraction of the maximum-mass star, which we quantify through its radius. We define the radius variation with respect to the most deleptonized configuration of this sequence, $Y_L^{\rm fin}=0.1$ at $S/A=1$, as
\begin{equation}
\label{eq:delta_Rmax}
   \Delta R_{\max}(Y_L) = R_{\max}(Y_L) - R_{\max}(Y_L^{\rm fin}) .
\end{equation}
In \cref{fig:delta_Rmax_YL} we show that the maximum-mass star contracts during deleptonization in most cases for both families of models.

The contraction is, however, considerably larger for most of the hyperonic EoSs, with $\Delta R_{\max}$ reaching about $1.6$~km, compared with at most $0.9$~km for the nucleonic ones. To identify the origin of this difference, we have separated the contraction of the core, defined as the region where $\rho_B > 0.08~\mathrm{fm}^{-3}$, from that of the outer layers\footnote{We have verified that the conclusions are unchanged if the core is defined by $\rho_B > 0.16~\mathrm{fm}^{-3}$.}. The outer layers contract by $0.5$ to $1.2$~km for every model, irrespective of its composition, as a result of the loss of the lepton pressure, whose relative contribution is largest at low densities. The two families differ in the behavior of the core. For the nucleonic models, the core does not contract and in most cases expands by up to $0.9$~km, consistent with the stiffening of the dense matter discussed above. For most of the hyperonic models, on the contrary, the core shrinks by $0.1$ to $0.6$~km, which is the structural counterpart of the softening produced by the hyperons, leading to the increase of the central density shown in \cref{fig:central_density_YL}. The exceptions are DNS and QMC-A, which have a low hyperon fraction, the purely kaonic DD2K, and STOS(TM1)Y. For the latter, the central density decreases below $Y_L \approx 0.25$ and the radius grows below $Y_L \approx 0.2$, but the same behavior is found for its nucleonic counterpart STOS(TM1), so that it originates from the underlying TM1 interaction rather than from the hyperons. In this model, the hyperons remove the gain in maximum baryonic mass that the deleptonization produces in the nucleonic case, without compressing the core. DNS, QMC-A and STOS(TM1)Y are also the only hyperonic models that display a slight, temporary expansion of the star at intermediate lepton fractions, visible as negative values of $\Delta R_{\max}$ in \cref{fig:delta_Rmax_YL}, a feature shared by the nucleonic models STOS(TM1) and LPB. In the latter, the stiffening produced by the deleptonization is so strong that the star does not contract at all, and its radius at $Y_L = 0.1$ is slightly larger than at $Y_L = 0.4$.

Finally, in order to characterize the stiffening or softening of the matter along the deleptonization path, we examine the evolution of an effective thermal index, computed with respect to the cold, $\beta$-equilibrated EoS. For a given thermodynamic point, the local effective thermal index is defined as
\begin{equation}
\label{eq:thermal_index}
\Gamma_{\rm th} = 1 + \frac{P(T, Y_L, \rho_B) - P_{0,\rm eq}}{\epsilon(T, Y_L, \rho_B) - \epsilon_{0,\rm eq}} ,
\end{equation}
where $P$ and $\epsilon$ are the pressure and the energy density of the finite-temperature, lepton-rich matter, while $P_{0,\rm eq}$ and $\epsilon_{0,\rm eq}$ denote the corresponding quantities of the cold, catalyzed matter evaluated at the same baryon density. Since the reference state is the cold, catalyzed matter and not cold matter with the same composition, the difference $P - P_{0,\rm eq}$ contains, besides the thermal contribution, the change of the pressure due to the different lepton and particle content, and for this reason we refer to $\Gamma_{\rm th}$ as an effective thermal index. We report its mass-average $\langle \Gamma_{\rm th} \rangle$ over the maximum-mass star, following the definition of \cref{eq:mass_average}. In its standard form, defined with respect to cold matter of the same composition, the thermal index has been widely employed as an indicator of the thermal stiffness of the EoS, in particular in the context of NS mergers~\cite{RadutaEoSHotNSExoticParticleDoF2022, BlackerThermalBehaviorIndicatorHyperonsBNSMergers2024, KochankovskiImpactHyperonsNSMergers2025}, and here it serves the analogous purpose of tracking whether the matter is getting softer or stiffer along the evolution.

\begin{figure*}[tp]
    \centering
    \subfigure[Hyperonic EoSs]{\includegraphics[width=0.45\textwidth]{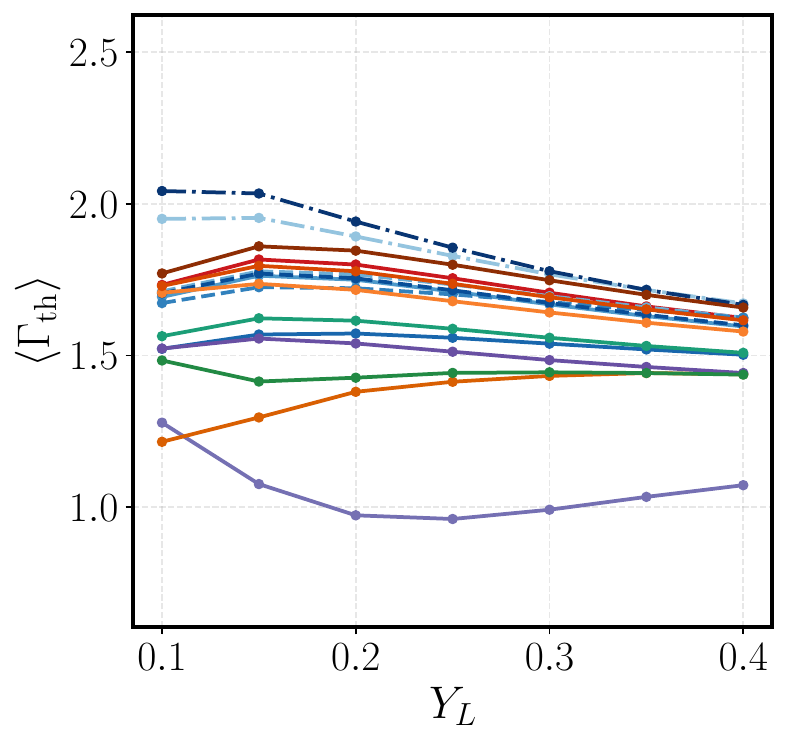}}
    \hfill
    \subfigure[Nucleonic EoSs]{\includegraphics[width=0.45\textwidth]{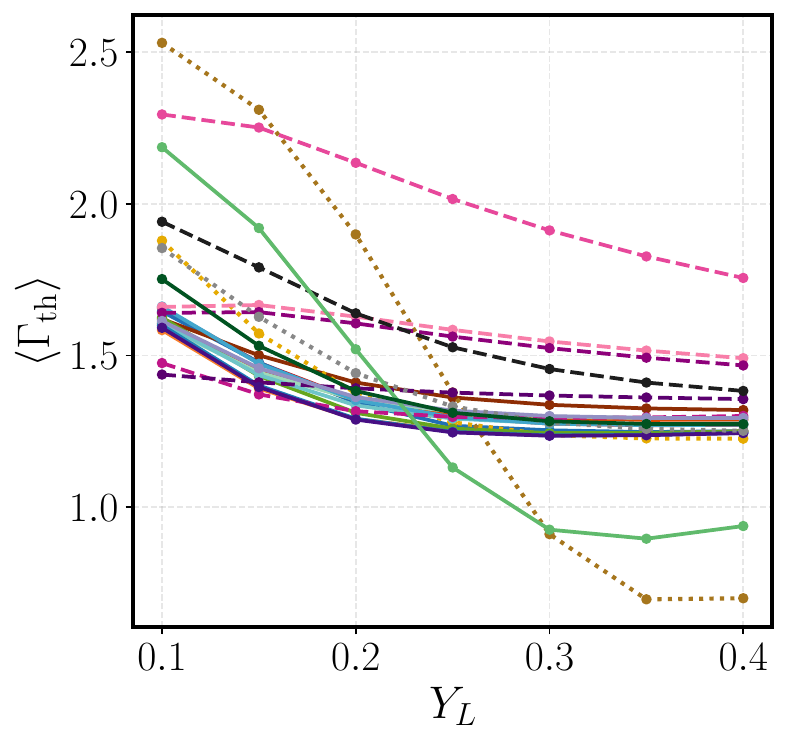}}
    \caption{Mass-averaged effective thermal index $\langle \Gamma_{\rm th} \rangle$,
    defined in \cref{eq:thermal_index}, as a function of the total lepton
    fraction $Y_L$, at fixed entropy per baryon $S/A = 1$, for the hyperonic
    EoSs (left panel) and the nucleonic EoSs (right panel). The line styles and
    colors are the same as in \cref{fig:cold_MR}.}
    \label{fig:thermal_index_YL}
\end{figure*}

The behavior of the average effective thermal index along the deleptonization path is shown in \cref{fig:thermal_index_YL}. For the nucleonic models, shown in the right panel, $\langle \Gamma_{\rm th} \rangle$ increases monotonically as the lepton fraction decreases, signaling the progressive stiffening of the EoS as the matter becomes more isospin-asymmetric. This also holds for the RB Skyrme models, some of which have been constrained to reproduce the soft thermal response predicted by chiral effective field theory. For two of them, RB2 and RB7, the effective thermal index levels off below $Y_L \approx 0.15$, but it does not decrease appreciably. For most of the hyperonic models, shown in the left panel, the effective thermal index initially increases as well, but it saturates around $Y_L \approx 0.15$ and then decreases. At this point, the softening produced by the hyperons outweighs the stiffening provided by the isospin asymmetry, and the EoS becomes softer as the star continues to deleptonize. Such a turnover is not found for any of the nucleonic models, and it is therefore a direct signature of the presence of hyperons. Two hyperonic models depart from this pattern, and both are consistent with the results of \cref{sec:PNSevolution}. For QMC-A, the effective thermal index decreases over the whole range of lepton fractions, from $1.44$ at $Y_L = 0.4$ to $1.21$ at $Y_L = 0.1$, which explains why even its small hyperon content is sufficient to trigger the delayed collapse. For DNS, on the contrary, the effective thermal index decreases only down to $Y_L \approx 0.25$ and increases again for lower lepton fractions, so that the stiffening produced by the deleptonization prevails over the softening produced by the hyperons, in line with the fact that this model does not predict a delayed collapse. The absolute values of the effective thermal index are not significant for the present discussion. What matters is whether it increases or decreases along the evolution, that is, whether the EoS is getting stiffer or softer.

\subsection{Heating and cooling}
\label{sec:heating_cooling}

Having isolated the effects of the deleptonization, we now turn to the thermal evolution and analyze its influence on the most relevant quantities characterizing the maximum-mass configurations. To this end, we fix the total lepton fraction to $Y_L = 0.2$ and let the entropy per baryon vary in the range $0 \leq S/A \leq 2$, thus covering the thermal conditions encountered by the star from the cold limit up to the hot intermediate stage of the evolution\footnote{As before, although we present the results obtained only at $Y_L = 0.2$, we have verified that the qualitative behavior is unchanged at different lepton fractions.}.

\begin{figure}[tbp]
    \centering
    \includegraphics[width=0.99\columnwidth]{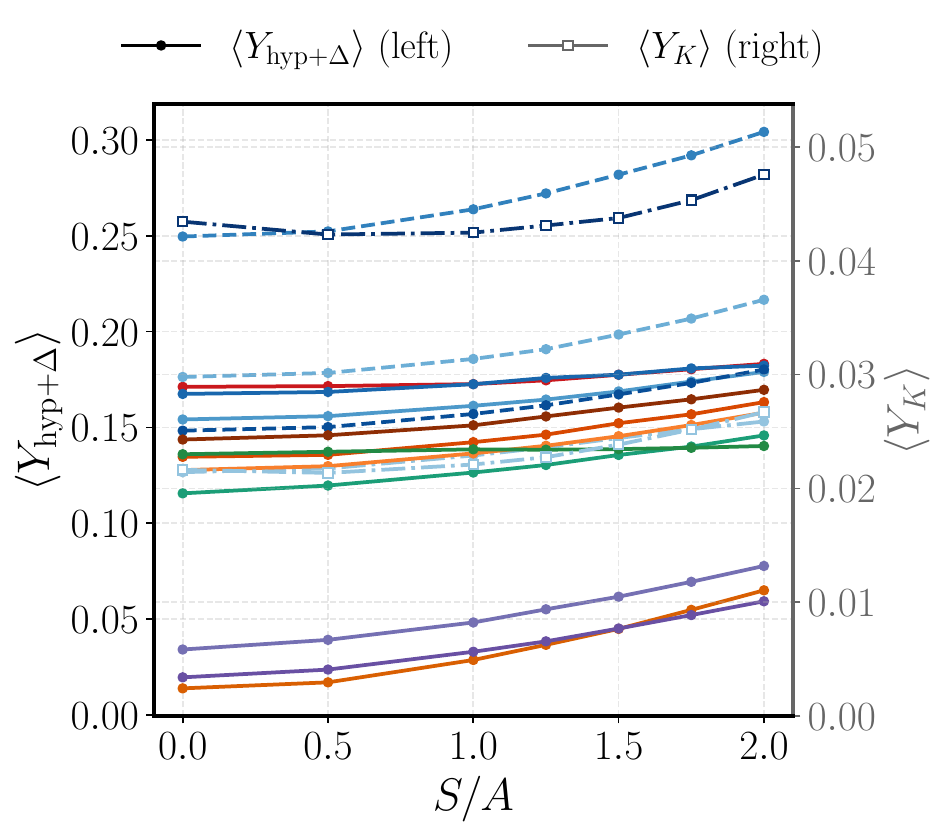}
    \caption{Same as \cref{fig:hyp_frac_YL}, but as a function of the entropy per
    baryon $S/A$ at fixed total lepton fraction $Y_L = 0.2$.}
    \label{fig:hyp_frac_S}
\end{figure}

As before, we start by showing in \cref{fig:hyp_frac_S} the mass-averaged hyperon fraction and the mass-averaged kaon fraction of the maximum-mass star as a function of the entropy per baryon at fixed $Y_L = 0.2$. As expected, for all the models, the hyperon content grows with increasing entropy, reflecting the thermal population of the hyperons and $\Delta$ resonances, which appear already below their threshold density in cold matter~\cite{MarquesTemperatureDependentHEoS2014, FortinHyperonsHotDenseMatter2018, KochankovskiEoSHotHyperonicNSM2022}. The effect size, however, depends strongly on the model. For most of the EoSs, $\langle Y_{\mathrm{hyp+\Delta}} \rangle$ increases by only 7 to 26 percent between the cold limit and $S/A = 2$, and for STOS(TM1)Y it is almost independent of the entropy. The thermal variation of the hyperon content therefore remains modest when compared to the change induced by the deleptonization, displayed in \cref{fig:hyp_frac_YL}. For DD2Y$\Delta$(1.2,1.1), for instance, $\langle Y_{\mathrm{hyp+\Delta}} \rangle$ rises from 0.17 to 0.32 as $Y_L$ decreases from 0.4 to 0.1, whereas heating from $S/A = 0$ to $S/A = 2$ raises it only from 0.25 to 0.30. The composition of the matter is therefore shaped mainly by the deleptonization, consistent with the picture drawn in \cref{sec:deleptonization}, in which the nucleonic and hyperonic families are distinguished by their response to the loss of leptons. The exceptions are SFHoY, DNS and QMC-A, whose small hyperon content grows by a factor of 2 to 5 with the entropy. This relative increase is comparable to the one produced by the deleptonization for SFHoY and smaller for DNS, and it exceeds it only for QMC-A, for which the thermal population of the hyperons is decisive, since the instability appears only at the hot intermediate stage, as discussed in \cref{sec:PNSevolution}. The total kaon fraction, which includes both condensed and thermal kaons, changes only mildly with the entropy, since the heating suppresses the condensate while it populates the thermal kaons, and the two effects nearly compensate each other.

\begin{figure*}[tp]
    \centering
    \subfigure[Hyperonic EoSs]{\includegraphics[width=0.45\textwidth]{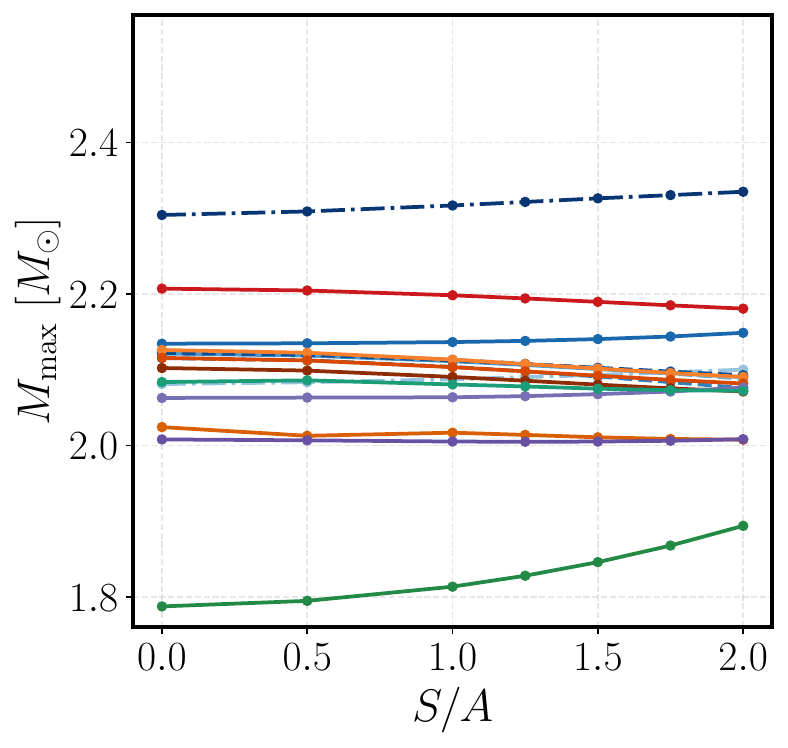}}
    \hfill
    \subfigure[Nucleonic EoSs]{\includegraphics[width=0.45\textwidth]{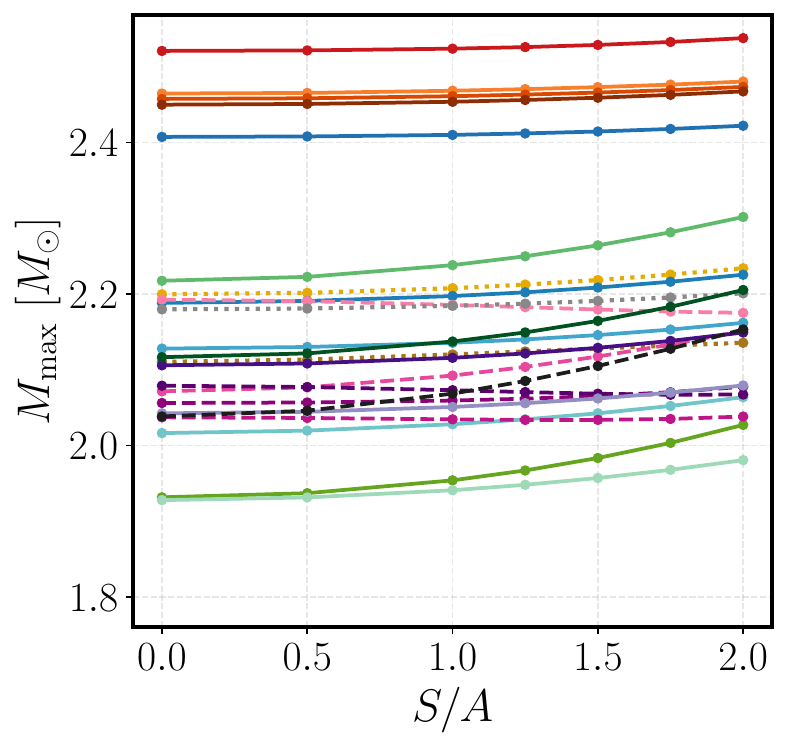}}
    \caption{Maximum gravitational mass $M_{\mathrm{max}}$ as a function of the
    entropy per baryon $S/A$, at fixed total lepton fraction $Y_L = 0.2$, for
    the hyperonic EoSs (left panel) and the nucleonic EoSs (right panel). The
    line styles and colors are the same as in \cref{fig:cold_MR}.}
    \label{fig:grav_mass_S}
\end{figure*}

In \cref{fig:grav_mass_S} we show the behavior of the maximum gravitational mass for the hyperonic EoSs (left panel) and for the nucleonic EoSs (right panel). For the hyperonic models, and contrary to what was found in the case of deleptonization, the change in entropy has only a mild effect on $M_{\mathrm{max}}$, and the direction of the effect is model-dependent. For most of the models with a large hyperon content, the maximum mass decreases with increasing entropy, while for the models with a small hyperon content, as well as for the kaonic ones, it increases slightly. The decrease in the maximum mass reflects the softening produced by the thermally populated hyperons, which will be discussed below in terms of the effective thermal index. The nucleonic EoSs, on the other hand, show a more consistent behavior, and for almost all of them the maximum gravitational mass increases with the entropy. The only exceptions are the Skyrme models RB2 and RB8, for which $M_{\mathrm{max}}$ decreases slightly, while for RB6 it remains essentially constant.

\begin{figure*}[tp]
    \centering
    \subfigure[Hyperonic EoSs]{\includegraphics[width=0.45\textwidth]{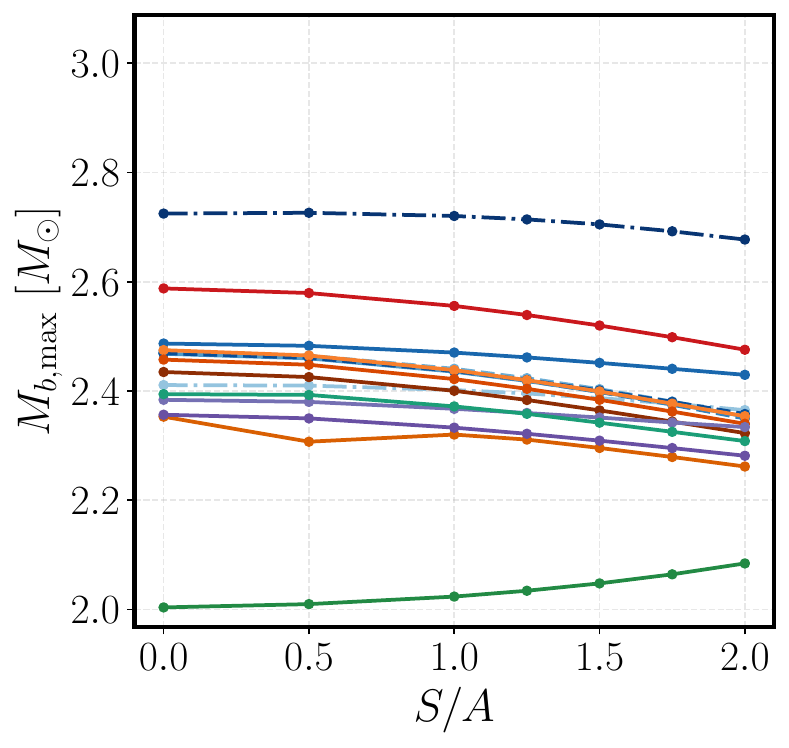}}
    \hfill
    \subfigure[Nucleonic EoSs]{\includegraphics[width=0.45\textwidth]{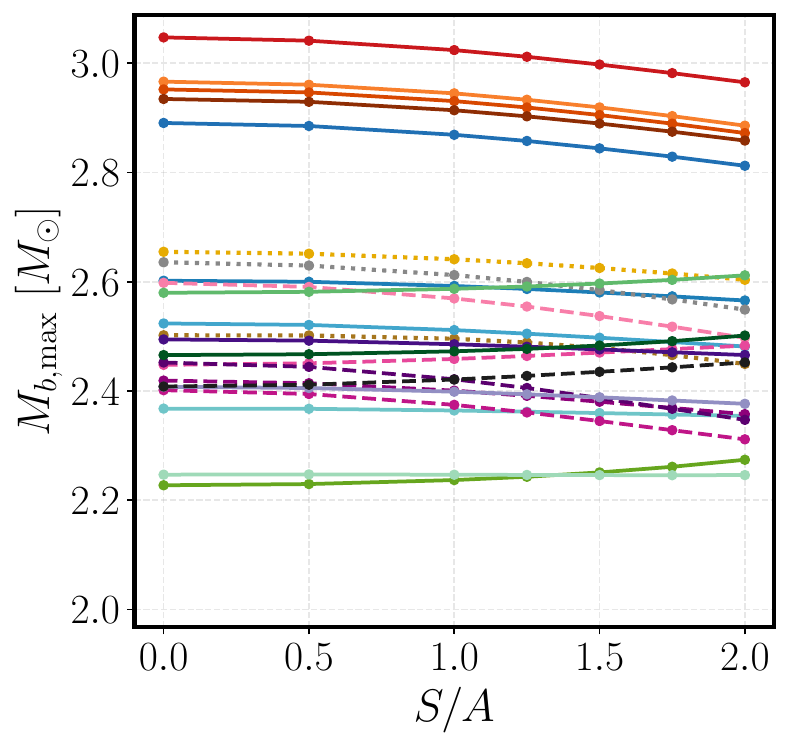}}
    \caption{Maximum baryonic mass $M_{b,\mathrm{max}}$ as a function of the
    entropy per baryon $S/A$, at fixed total lepton fraction $Y_L = 0.2$, for
    the hyperonic EoSs (left panel) and the nucleonic EoSs (right panel). The
    line styles and colors are the same as in \cref{fig:cold_MR}.}
    \label{fig:bar_mass_S}
\end{figure*}

In \cref{fig:bar_mass_S} we display the maximum baryonic mass as a function of the entropy per baryon. For most of the EoSs, it decreases monotonically with the entropy, so that, contrary to the deleptonization case, it does not follow the trend of $M_{\mathrm{max}}$. Between the cold limit and $S/A = 2$ the reduction amounts to $0.05$ to $0.12\,M_\odot$ for the hyperonic models and up to $0.11\,M_\odot$ for the nucleonic ones. The exceptions are the nucleonic models IUF, RB5, SRO(SLy4), STOS(TM1) and TM1e, together with the hyperonic STOS(TM1)Y, for which $M_{b,\mathrm{max}}$ increases slightly with the entropy. The heating therefore affects the two families in the same direction and by a comparable amount, somewhat more strongly when hyperons are present. It cannot by itself account for the different fate of the hyperonic and nucleonic stars.

\begin{figure*}[tp]
    \centering
    \subfigure[Hyperonic EoSs]{\includegraphics[width=0.45\textwidth]{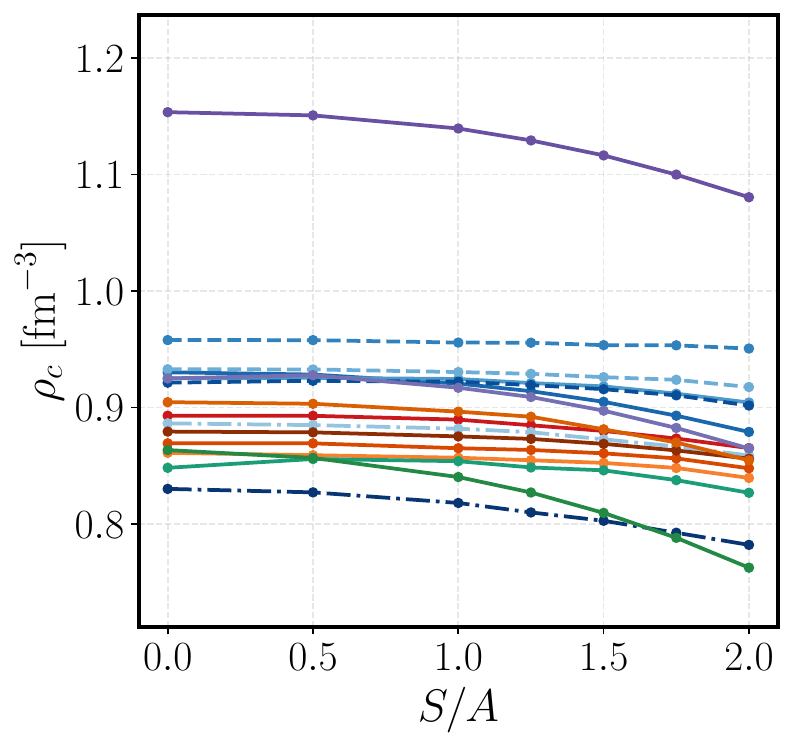}}
    \hfill
    \subfigure[Nucleonic EoSs]{\includegraphics[width=0.45\textwidth]{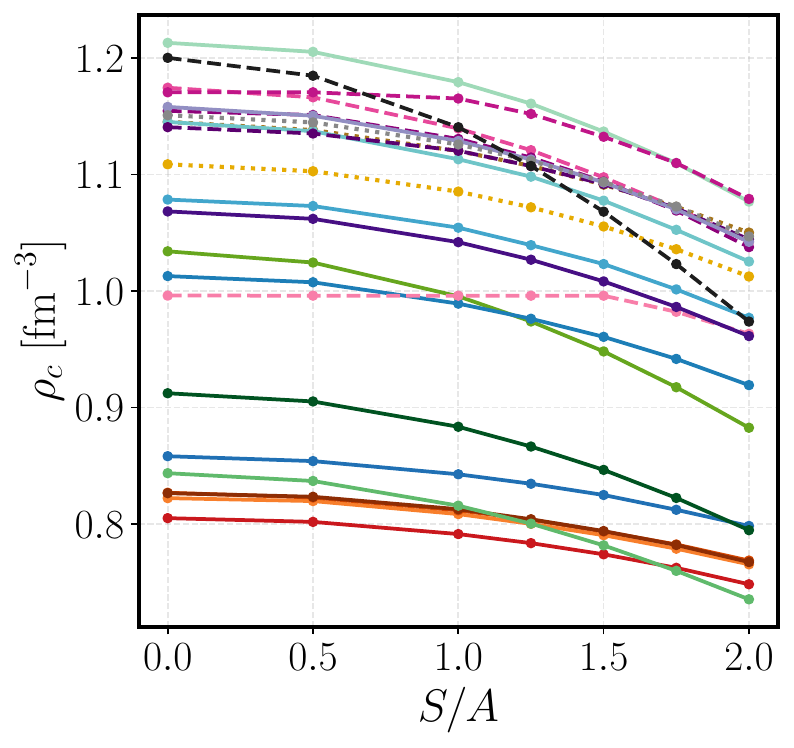}}
    \caption{Central density $\rho_c$ of the maximum-mass star as a function of
    the entropy per baryon $S/A$, at fixed total lepton fraction $Y_L = 0.2$,
    for the hyperonic EoSs (left panel) and the nucleonic EoSs (right panel).
    The line styles and colors are the same as in \cref{fig:cold_MR}.}
    \label{fig:central_density_S}
\end{figure*}

We now examine how the central density of the maximum-mass configuration evolves with the entropy per baryon at fixed lepton fraction, shown in \cref{fig:central_density_S}. For both nucleonic and hyperonic families, the central density decreases as the entropy increases, so that the star is less compact when it is hot and contracts when it cools. Contrary to the deleptonization, where only the hyperonic models showed a systematic trend, as can be appreciated by comparing with \cref{fig:central_density_YL}, this thermal response is thus a generic feature of both classes of EoSs. The size of the effect, however, is different. Between $S/A = 0$ and $S/A = 2$ the central density decreases by $0.01$ to $0.07\,\mathrm{fm}^{-3}$ for most of the hyperonic models and by $0.03$ to $0.23\,\mathrm{fm}^{-3}$ for the nucleonic ones. For the DD2 interaction, for instance, the reduction amounts to $0.06\,\mathrm{fm}^{-3}$ in the nucleonic case and to only $0.02\,\mathrm{fm}^{-3}$ when hyperons are included, since part of the additional thermal pressure is compensated by the softening produced by the thermally populated hyperons.

\begin{figure*}[tp]
    \centering
    \subfigure[Hyperonic EoSs]{\includegraphics[width=0.45\textwidth]{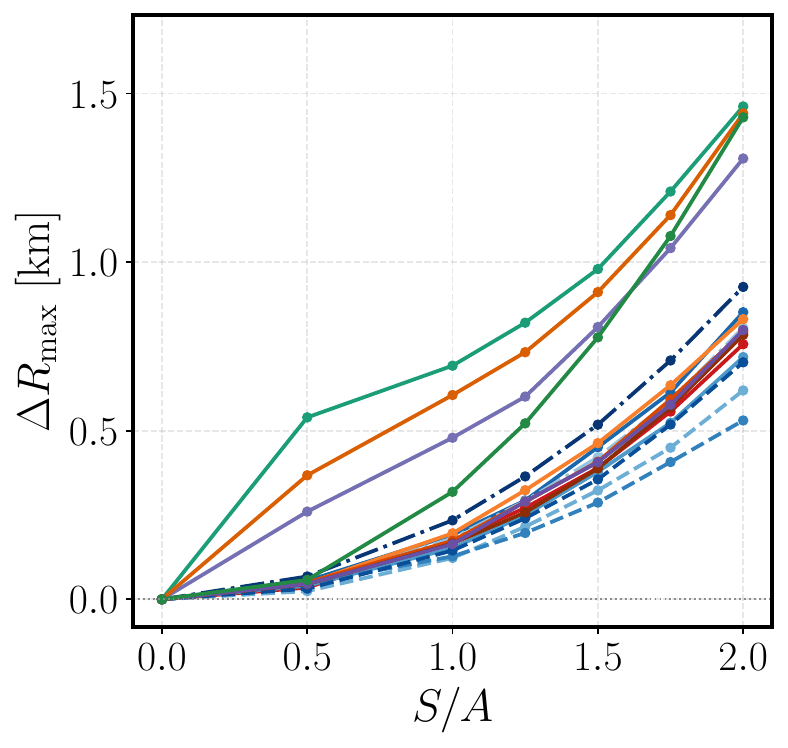}}
    \hfill
    \subfigure[Nucleonic EoSs]{\includegraphics[width=0.45\textwidth]{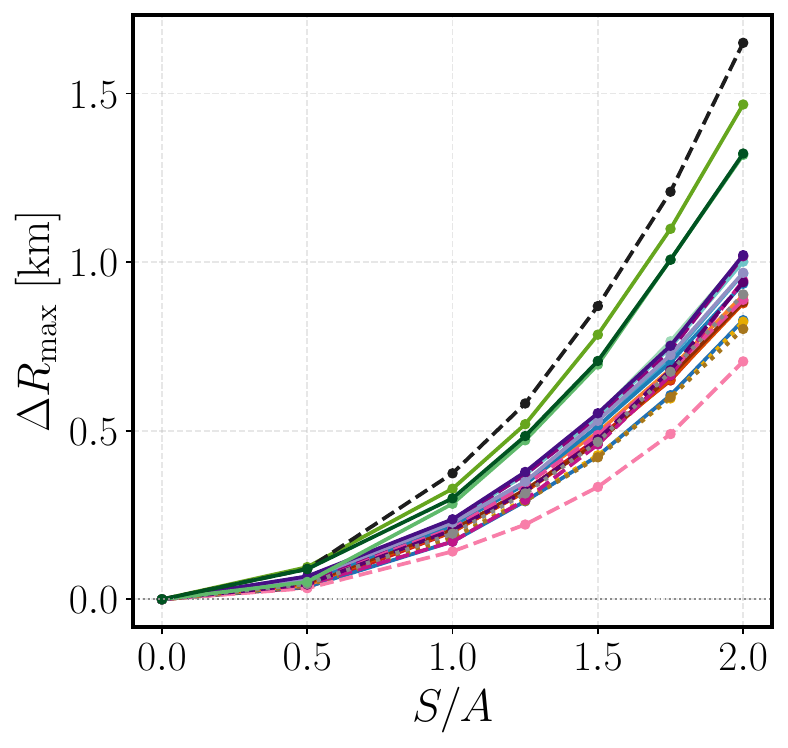}}
    \caption{Variation of the radius of the maximum-mass star,
    $\Delta R_{\max}$, defined in \cref{eq:delta_Rmax_S}, as a function of
    the entropy per baryon $S/A$, at fixed total lepton fraction $Y_L = 0.2$,
    for the hyperonic EoSs (left panel) and the nucleonic EoSs (right panel).
    The line styles and colors are the same as in \cref{fig:cold_MR}.}
    \label{fig:delta_Rmax_S}
\end{figure*}

The evolution of the radius of the maximum-mass star with the entropy per baryon at fixed lepton fraction is shown in \cref{fig:delta_Rmax_S}. In analogy with \cref{eq:delta_Rmax}, we quantify it through the radius variation measured with respect to the cold configuration,
\begin{equation}
\label{eq:delta_Rmax_S}
   \Delta R_{\max}(S/A) = R_{\max}(S/A) - R_{\max}(S/A = 0) .
\end{equation}
The behavior is qualitatively similar for both nucleonic and hyperonic families, since the star expands monotonically with increasing entropy, as expected from the additional thermal pressure support\footnote{For FSU2H$^{*}$, QMC-A and DNS the temperature grids of the tables are coarse at low temperatures, with the first nonzero temperature at $10$, $10$ and $2$~MeV, respectively. At $S/A = 0.5$ the temperature of the outer layers of the star falls below this value, where the interpolation overestimates the thermal pressure. The sizable expansion of these three models between $S/A = 0$ and $S/A = 0.5$ is therefore a numerical artifact, and the same applies to the first entropy step of the other quantities shown in this subsection for these models.}. Contrary to the deleptonization, where the hyperonic stars contracted considerably more, the magnitude of the expansion is here comparable between the two classes, with $\Delta R_{\max}$ at $S/A = 2$ between $0.5$ and $1.5$~km for the hyperonic models and between $0.7$ and $1.7$~km for the nucleonic ones.

\begin{figure*}[tp]
    \centering
    \subfigure[Hyperonic EoSs]{\includegraphics[width=0.45\textwidth]{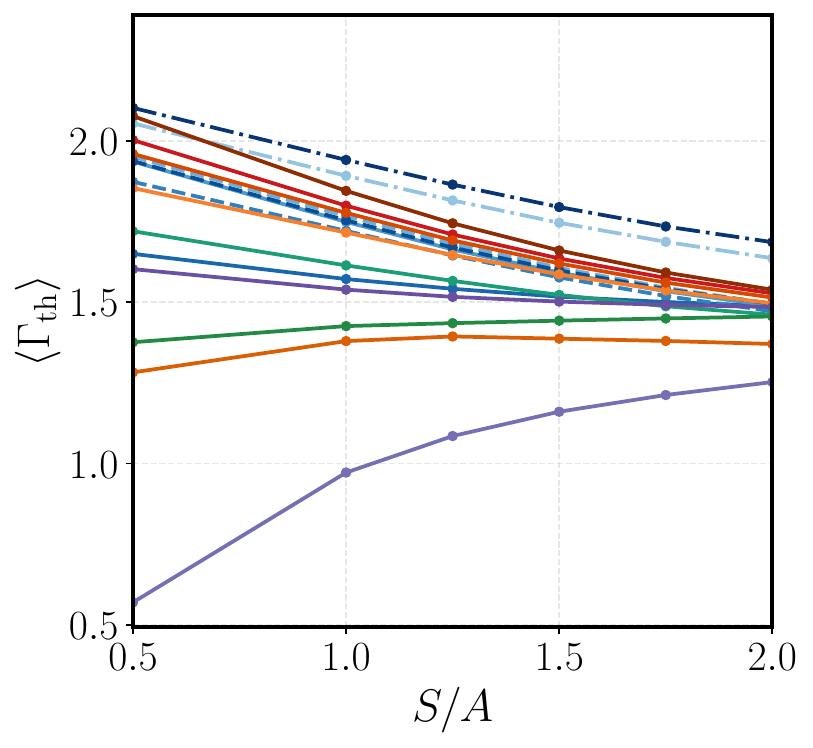}}
    \hfill
    \subfigure[Nucleonic EoSs]{\includegraphics[width=0.45\textwidth]{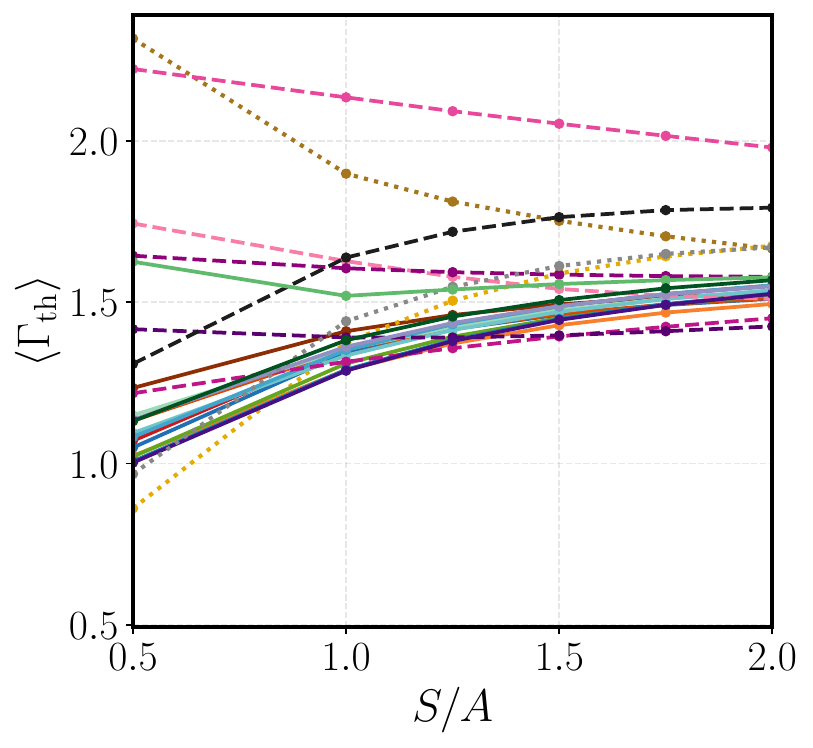}}
    \caption{Mass-averaged effective thermal index $\langle \Gamma_{\rm th} \rangle$,
    defined in \cref{eq:thermal_index}, as a function of the entropy per
    baryon $S/A$, at fixed total lepton fraction $Y_L = 0.2$, for the hyperonic
    EoSs (left panel) and the nucleonic EoSs (right panel). The line styles and
    colors are the same as in \cref{fig:cold_MR}.}
    \label{fig:thermal_index_S}
\end{figure*}

Finally, in \cref{fig:thermal_index_S} we show the mass-averaged effective thermal index $\langle \Gamma_{\rm th} \rangle$, defined in \cref{eq:thermal_index}, as a function of the entropy per baryon at fixed $Y_L = 0.2$. At low entropy, the thermal contribution to $\epsilon - \epsilon_{0,\rm eq}$ is small, so that the effective thermal index is dominated by the difference in composition between the lepton-rich matter and the cold, catalyzed reference, and its values at the lowest entropies have to be interpreted with care. For most of the hyperonic models, collected in the left panel, $\langle \Gamma_{\rm th} \rangle$ decreases with increasing entropy, since the thermal population of the hyperons grows with the temperature and progressively softens the EoS. The kaonic models DD2K and BHB$\Lambda\phi$K are among those with the highest effective thermal index over the whole range of entropies. The exceptions to the decreasing trend are DNS and STOS(TM1)Y, for which the effective thermal index increases, and QMC-A, for which it is non-monotonic.

The nucleonic models, shown in the right panel of \cref{fig:thermal_index_S}, behave differently, and for most of them the effective thermal index increases with the entropy, from values around $1$ to about $1.5$. Several nucleonic models, namely the microscopic LPB and the Skyrme models RB2, RB5 and RB7, nevertheless display a decreasing effective thermal index, like the hyperonic ones, while for RB8 and STOS(TM1) it is non-monotonic. The thermal index has been proposed as an indicator of exotic degrees of freedom in the context of NS mergers~\cite{RadutaEoSHotNSExoticParticleDoF2022, BlackerThermalBehaviorIndicatorHyperonsBNSMergers2024, KochankovskiImpactHyperonsNSMergers2025}. In the present context, however, the response to the heating alone does not separate the two families, since a nucleonic EoS with a soft thermal response can display the same trend as a hyperonic one. Nevertheless, as seen in Sec.~\ref{sec:PNSevolution}, for some hyperonic models the additional thermal softening triggers the delayed collapse.

Summarizing the results of the two subsections, the disentangling of the two processes shows that the nucleonic and hyperonic EoSs differ in how they respond to deleptonization, and not to heating. The heating lowers the maximum baryonic mass of both families by a comparable amount. The deleptonization, on the contrary, raises $M_{b,\mathrm{max}}$ by $0.07$ to $0.31\,M_\odot$ for every nucleonic EoS, more than enough to compensate the thermal loss, while for most of the hyperonic EoSs it leaves $M_{b,\mathrm{max}}$ almost unchanged or lowers it, so that the thermal loss is not compensated. In the absence of exotic hadrons, deleptonization always stiffens the matter, and it is the suppression of this stiffening by the hyperons that drives the most massive PNSs described by hyperonic EoSs towards the delayed collapse.

\section{Thermal and structural imprints of exotic hadrons on the maximum-mass star}
\label{sec:other_results}

We now examine how exotic hadrons modify the temperature, the central density, and the radius of the maximum-mass star, which is the configuration in which they are most abundant. While in \cref{sec:cooling_deleptonization} the deleptonization and the thermal evolution were isolated one at a time to identify the dominant driver of the instability, here we adopt a complementary point of view and analyze the physical snapshots of the evolution, collecting the whole set of models at the early and intermediate stages. The latter is the stage at which the hyperon content is the largest for most of the models and at which the instability sets in for most of the hyperonic EoSs, as discussed in \cref{sec:PNSevolution}.

\begin{figure*}[tp]
    \centering
    \subfigure[]{\includegraphics[width=0.48\textwidth]{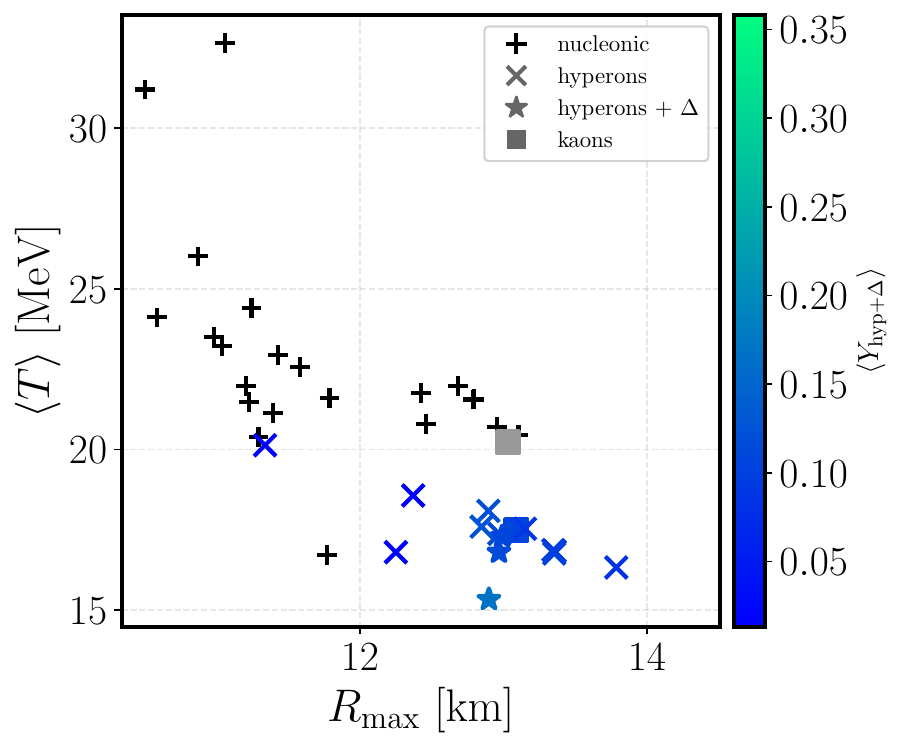}}
    \subfigure[]{\includegraphics[width=0.48\textwidth]{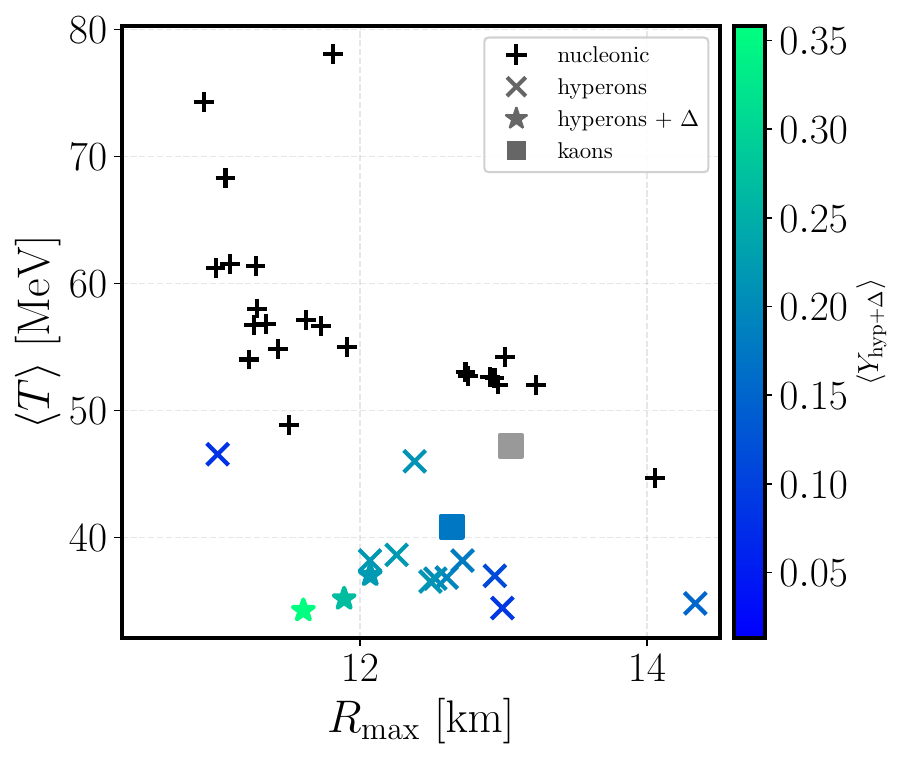}}
    \caption{Mass-averaged temperature $\langle T \rangle$ of the maximum-mass stars
    as a function of their radius $R_{\mathrm{max}}$, during the early stage
    ($Y_L = 0.4$, $S/A = 1$, left panel) and the intermediate stage
    ($Y_L = 0.1$, $S/A = 2$, right panel). The marker style distinguishes the
    composition of the model (nucleonic, hyperonic, hyperonic with $\Delta$
    resonances, and models with a kaon condensate), while the color encodes the
    mass-averaged hyperon (plus $\Delta$) fraction $\langle Y_{\mathrm{hyp+\Delta}} \rangle$.}
    \label{fig:avgT_vs_R}
\end{figure*}

Besides the pressure of matter, reflected in the effective thermal index discussed in \cref{sec:cooling_deleptonization}, the hyperons also modify the temperature reached at a given entropy per baryon. In \cref{fig:avgT_vs_R} we show the mass-averaged temperature of the maximum-mass stars during the early (left panel) and the intermediate (right panel) stages as a function of their radius. The color encodes the mass-averaged hyperon fraction in the corresponding configuration, while the marker style indicates the composition of the model. The results of the nucleonic models follow a correlation between the temperature and the radius, with smaller radii reaching higher temperatures, although the scatter is considerable. At fixed entropy per baryon, a more compact configuration reaches higher densities in its interior and is therefore hotter. The temperatures for the hyperonic models lie systematically below the nucleonic ones already during the early phase, even though the color bar indicates a low hyperon content for almost all of them. The separation becomes considerably larger in the intermediate phase, once the lepton fraction is reduced and the hyperons populate the medium in significant amounts. The comparison between models that share the same nucleonic interaction allows us to isolate the effect of hyperons. For the BCP, DD2, DDLS, and STOS(TM1) families, the inclusion of hyperons lowers the mass-averaged temperature by 16\% to 22\%  in the early stage and by 22\% to 30\%  in the intermediate one, while for SFHo, whose hyperonic version carries the smallest hyperon content, the reduction amounts to 8\% and 18\% in the early and the intermediate phase, respectively. The DD2 interaction is also the nucleonic counterpart of BHB$\Lambda\phi$ and of the three DD2Y$\Delta$ models. In the intermediate stage, the reduction of $\langle T \rangle$ for these models with respect to DD2 amounts to 13\% for BHB$\Lambda\phi$, which contains only the $\Lambda$ hyperon, to 28\% for DD2Y and to 30\% to 35\%  for the DD2Y$\Delta$ models, although BHB$\Lambda\phi$, DD2Y and DD2Y$\Delta$(1.2,1.3) carry a similar hyperon fraction. The reason is that, for a fixed entropy per baryon, the presence of additional degrees of freedom allows the entropy to be distributed among a larger number of particle species, which lowers the temperature of the matter, as was shown in Refs.~\cite{Drago:2016srv, RadutaEoSHotNSExoticParticleDoF2022, KochankovskiEoSHotHyperonicNSM2022}. The same mechanism is responsible for the different heating of the nucleonic and hyperonic models along the deleptonization sequence of \cref{sec:deleptonization}, in which the entropy per baryon is kept fixed at $S/A = 1$, and only the lepton fraction is reduced from $Y_L = 0.4$ to $Y_L = 0.1$. Along this sequence, the entropy carried by the escaping neutrinos is transferred to the matter, and $\langle T \rangle$ increases by about $6$ to $8$~MeV for most of the nucleonic models, while for most of the hyperonic ones, where this entropy is shared with the newly populated hyperons, the increase amounts to only $1$ to $2$~MeV. 

\begin{figure*}[tp]
    \centering
    \subfigure[]{\includegraphics[width=0.48\textwidth]{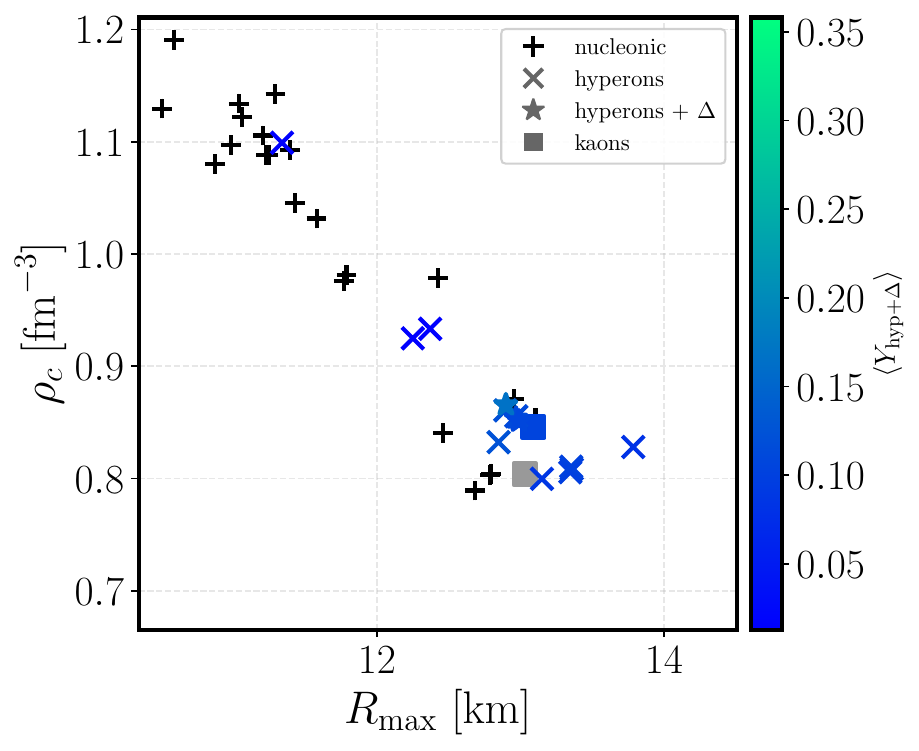}}
    \subfigure[]{\includegraphics[width=0.48\textwidth]{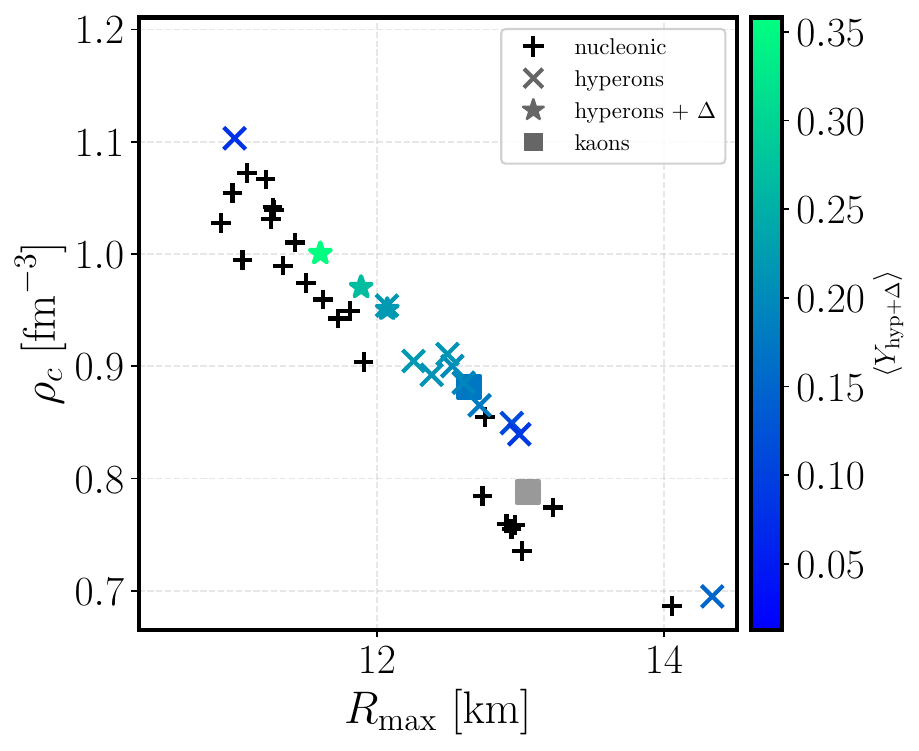}}
    \caption{Central baryon number density $\rho_c$ of the maximum-mass stars as a
    function of their radius $R_{\mathrm{max}}$, during the early stage
    ($Y_L = 0.4$, $S/A = 1$, left panel) and the intermediate stage
    ($Y_L = 0.1$, $S/A = 2$, right panel). The marker style and the color have the
    same meaning as in \cref{fig:avgT_vs_R}.}
    \label{fig:rhoc_vs_R}
\end{figure*}

A second correlation appears when we compare the central density of the maximum-mass stars with their radius. In \cref{fig:rhoc_vs_R}, this comparison is shown for the early (left panel) and the intermediate (right panel) phase. As expected, there is a tight correlation between the two quantities, since a smaller radius, signaling a softer EoS, corresponds to a higher central density. During the early phase, this relation is essentially composition-independent, with the hyperonic and the nucleonic models lying on top of each other. In the intermediate phase, however, once the hyperons are abundant, the hyperonic EoSs tend to deviate from the nucleonic relation and to lie slightly above it, displaying a somewhat higher central density for a fixed radius. This deviation, most pronounced for the models that also include $\Delta$ resonances,  indicates the increase of compactness due to the presence of new species, as discussed in \cref{sec:deleptonization}.

\begin{figure}[tbp]
    \centering
    \includegraphics[width=\columnwidth]{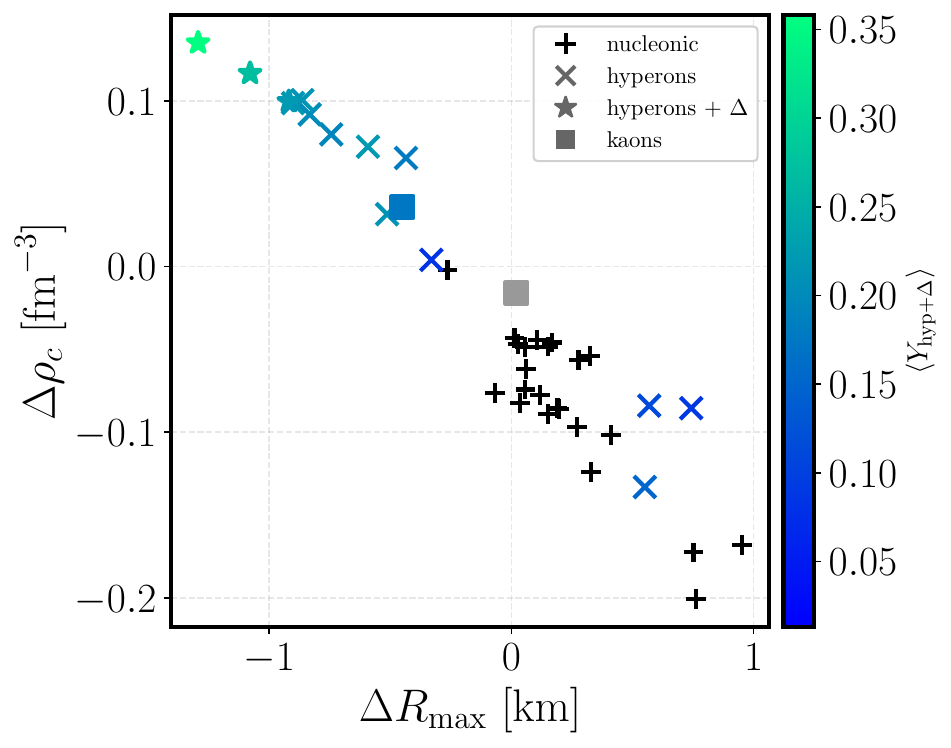}
    \caption{Change of the central density $\Delta \rho_c$ as a function of the
    change of the radius $\Delta R_{\mathrm{max}}$ of the maximum-mass stars
    between the intermediate and the early stage (intermediate minus early). The
    marker style and the color have the same meaning as in \cref{fig:avgT_vs_R}.}
    \label{fig:delta_rhoc_vs_delta_R}
\end{figure}

 We now plot the change of the central density between the two stages against the change of the radius, as shown in \cref{fig:delta_rhoc_vs_delta_R}, where $\Delta \rho_c$ and $\Delta R_{\mathrm{max}}$ denote the differences of these two quantities between the intermediate and early stages. Since the intermediate stage differs from the early one both in its lepton fraction and in its entropy per baryon, this plane superposes the effects of deleptonization and the heating that were isolated in \cref{sec:cooling_deleptonization}. We find that the EoS models separate neatly according to their hyperon content. The hyperonic models with a significant amount of hyperons occupy the upper-left part of the plot, reaching the most extreme values for the models that include $\Delta$ resonances. For these configurations, the strangeness-driven (and resonance-driven)  softening produced by the deleptonization dominates over the thermal expansion of the hotter intermediate stage, so that the star contracts and its central density grows. The nucleonic models, on the contrary, are found in the lower-right part of the plot, since the deleptonization induces a smaller contraction than for the hyperonic models and leaves the central density essentially unchanged. Hence, the net response is dictated by the thermal expansion of the higher-entropy stage, which enlarges the radius and lowers the central density. The separation is governed by the actual hyperon abundance, as the hyperonic EoSs that develop only a marginal hyperon content, namely DNS, QMC-A and SFHoY, behave like the nucleonic models and their results are found among the nucleonic ones or at the boundary between the two groups. The only exception to this picture is STOS(TM1)Y, which, despite a sizable hyperon fraction, lies in the lower-right region. The comparison with its nucleonic counterpart shows, however, that it behaves like the other hyperonic models. STOS(TM1) displays the largest expansion of the whole set, with $\Delta R_{\mathrm{max}} = 0.95$~km and $\Delta \rho_c = -0.17$~fm$^{-3}$, and the inclusion of hyperons reduces these values to $0.55$~km and $-0.13$~fm$^{-3}$, respectively.  The location of STOS(TM1)Y in the plot therefore originates from the underlying TM1 interaction, as was already found for the change of the radius along the deleptonization stage in \cref{sec:deleptonization}.

Overall, \cref{fig:delta_rhoc_vs_delta_R} condenses the competition between deleptonization and heating into a single diagnostic plane: a significant abundance of exotic hadrons alters the behavior of an evolving proto-neutron star from 
being dominated by thermal expansion to being dominated by deleptonization-driven contraction.
\section{Robustness of the nucleonic results: Chiral-EFT-constrained RMF models}

\label{sec:RMFNucleonic}

The nucleonic EoSs discussed so far differ not only in their nuclear matter properties, but also in the underlying many-body framework, in the treatment of the thermal effects, and in the density range in which they are constrained. While this heterogeneity makes the absence of a delayed collapse a rather general result, the sample is not a controlled scan of the nuclear matter parameters, since several of them change from one model to another. It remains to be checked whether our result survives a systematic variation of the parameters within a single nucleonic model.  

To address this question in a controlled way, we complement our analysis with a set of parametrizations within a single RMF model. Specifically, we use a finite-temperature extension~\cite{Christian:2023byx} of the EoS presented by Hornick et al.~\cite{HornickRelativisticParametrizationNeutronMatterNS2018} with parameter ranges that are compatible with chiral effective field theory ($\chi$EFT) at zero temperature and terrestrial nuclear data. In all parameter sets the saturation density, $\rho_0=0.150\,\mathrm{fm}^{-3}$, and the compressibility at $\rho_0$, $K_{\rm sat}=240$~MeV, are fixed. The three main variable parameters of this approach are the effective nucleon mass at $\rho_0$, which ranges from $m^{*}/m=0.55$ to $0.75$, the symmetry energy at $\rho_0$, $J=30$ and $32$~MeV, and its slope $L=40$--$60$~MeV. The quantities $m^{*}/m$ and $L$ have a much larger direct impact on the EoS around saturation density than $J$. Some combinations of $m^{*}/m$ and $L$ are, however, only acceptable for a specific value of $J$, which is why the symmetry energy is also slightly varied. In the following, we simply write $m^*$ for the ratio $m^{*}/m$.

In total we consider 39 $(m^*, L, J)$ parameter set combinations. The PNS snapshots are constructed exactly as for the other models of this work, i.e.~at constant total lepton fraction and entropy per baryon, including the contribution of trapped neutrinos\footnote{Below $\rho_B=0.01\,\mathrm{fm}^{-3}$ we employ the low-density part of the DD2 EoS~\cite{HempelStatisticalModelEoS2010, TypelCompositionThermodynamicsNuclearMatterLightClusters2010}, taken at the same entropy per baryon, or in cold $\beta$-equilibrium for the cold stars, and we connect it to the RMF table, used above $\rho_B=0.1\,\mathrm{fm}^{-3}$, through a log-log interpolation of the pressure and of the energy density. While the radius, especially for the low-mass and intermediate-mass stars, can be affected by up to a few tens of percent, $M_{\rm{max}}$ and $M_{\rm{b,max}}$ are not affected by the prescription.}.

\begin{figure}[t]
    \centering
    \includegraphics[width=0.90\columnwidth]{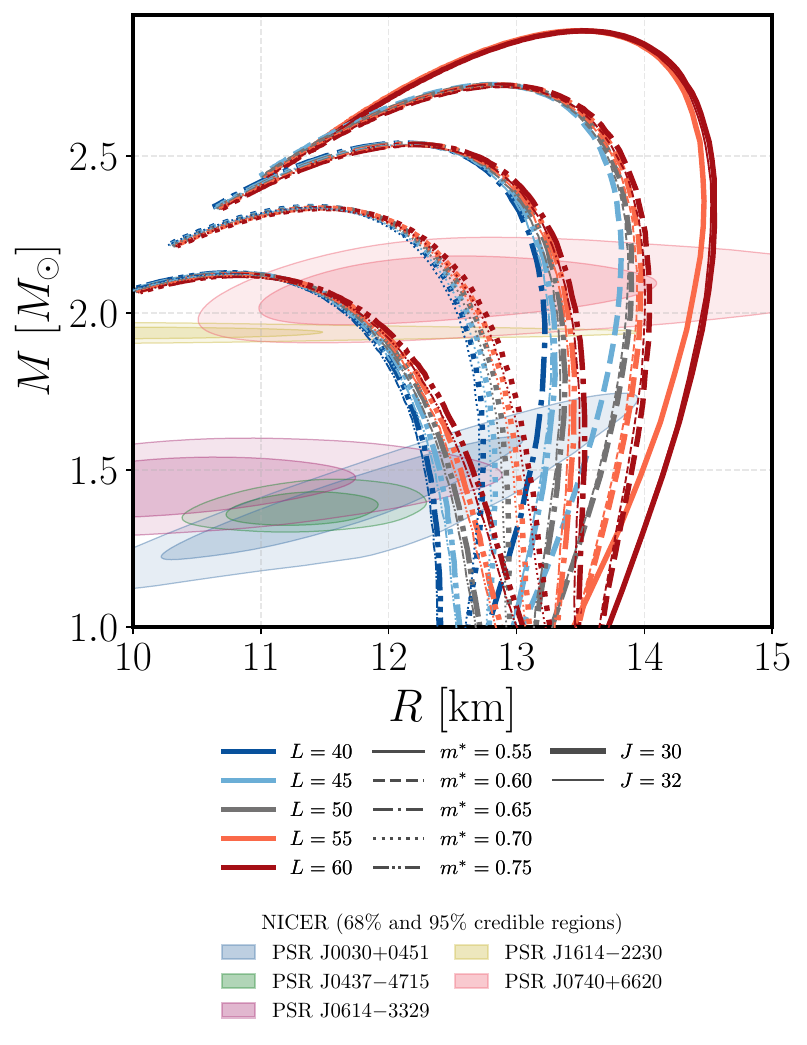}
    \caption{Mass-radius relations of cold, $\beta$-equilibrated NSs for the 39
    $\chi$EFT-constrained RMF parametrizations. The color indicates the slope
    of the symmetry energy $L$, the linestyle the effective mass $m^*$, and the
    line width the symmetry energy $J$ (thick: $J=30$~MeV, thin: $J=32$~MeV). The shaded areas show the 68\% and 95\% credible regions of the NICER mass-radius measurements of PSR J0030+0451~\cite{VinciguerraUpdatedMassRadiusAnalysisNICER2024}, PSR J0437$-$4715~\cite{Stammler:2026xaq}, PSR J0614$-$3329~\cite{MauviardANicerView14MSunPulsar2025}, PSR J1614$-$2230~\cite{MauviardNICERViewJ1614MassiveCompactMillisecondPulsar2026} and PSR J0740+6620~\cite{SalmiRadiusHighMassPulsarJ0740NICERData2024}.} 
    \label{fig:xeft_cold_MR}
\end{figure}

The mass-radius relations of the corresponding cold, $\beta$-equilibrated stars are shown in \cref{fig:xeft_cold_MR}. The curves are organized in five groups, one for each value of the effective mass: the stiffness of the EoS at high densities, and hence the maximum mass, is almost entirely determined by $m^*$. A lower effective mass implies a stronger scalar field and, consequently, a stronger vector repulsion required to reproduce the saturation properties, which results in a stiffer EoS at supra-saturation densities. The cold maximum mass therefore decreases from $M_{\max}\simeq 2.90\,M_\odot$ for $m^*=0.55$ to $M_{\max}\simeq 2.12\,M_\odot$ for $m^*=0.75$, while the radius of the maximum-mass star decreases from $\simeq 13.5$~km to $\simeq 10.8$~km. Within each group, $L$ and $J$ change $M_{\max}$ by less than $0.015\,M_\odot$. They only influence the radii of the intermediate-mass stars, which are governed by the density dependence of the symmetry energy around and slightly above saturation density: within a group of fixed $m^*$, the radius of the $1.4\,M_\odot$ star varies by up to $\approx 0.45$~km, with larger values of $L$ leading to larger radii. The effect of $J$ is even smaller: at fixed $m^*$ and $L$, increasing $J$ from $30$ to $32$~MeV shifts the maximum mass by only $\approx -0.005\,M_\odot$ and the radii by a few tens of meters.

\subsection{Maximum gravitational and baryonic masses}

\begin{figure*}[tp]
    \centering
    \includegraphics[width=0.45\textwidth]{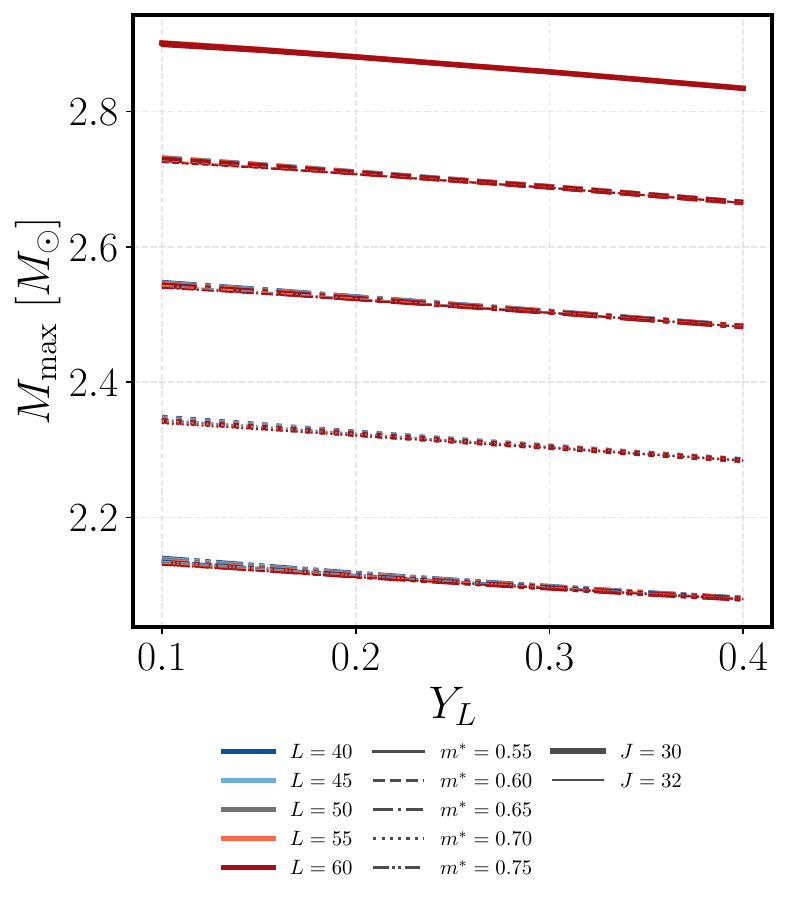}
    \hfill
    \includegraphics[width=0.45\textwidth]{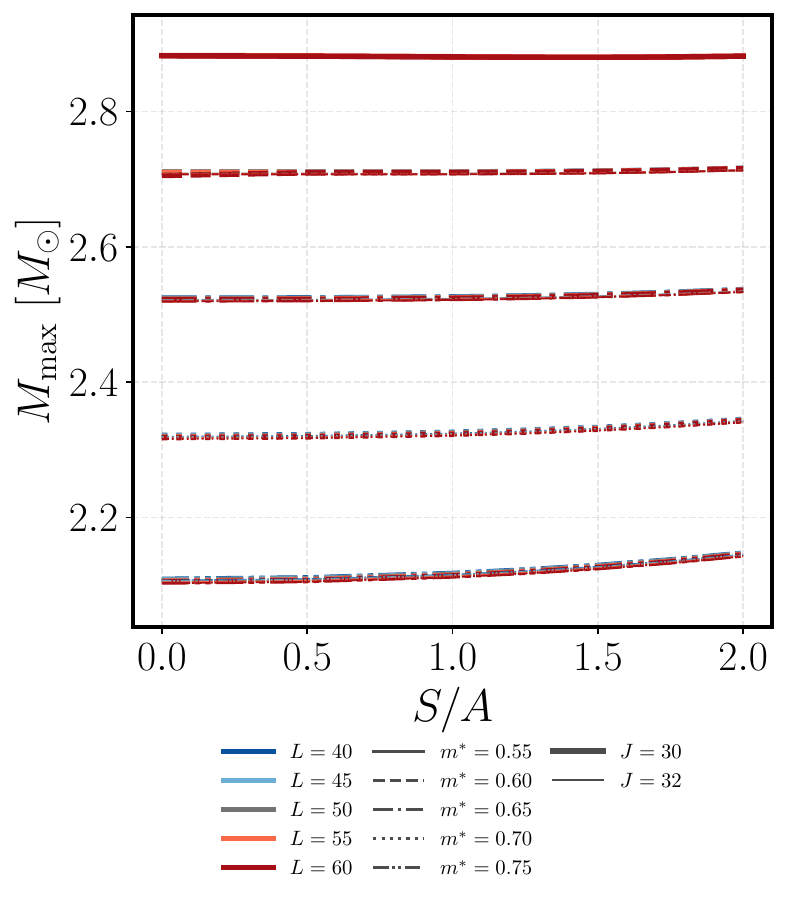}
    \caption{Maximum gravitational mass for the $\chi$EFT-constrained RMF parametrizations. Left: as a function of the lepton fraction $Y_L$ at constant $S/A=1$. Right: as a function of the entropy per baryon at constant $Y_L=0.2$. Colors, linestyles and line widths as in \cref{fig:xeft_cold_MR}.}
    \label{fig:xeft_Mmax}
\end{figure*}

As previously discussed, the maximum gravitational and baryonic masses determine the fate of the most massive stars during evolution. \cref{fig:xeft_Mmax} shows $M_{\max}$ as a function of the lepton fraction at constant entropy $S/A=1$ (left panel) and as a function of the entropy per baryon at constant lepton fraction $Y_L=0.2$ (right panel). As anticipated from the cold mass-radius relations, the magnitude of $M_{\max}$ is set by the effective mass, and the curves corresponding to different values of $L$ and $J$ lie essentially on top of each other. The deleptonization then acts in the same way for all the parametrizations: $M_{\max}$ increases monotonically with decreasing $Y_L$, by $0.05$--$0.07\,M_\odot$ between $Y_L=0.4$ and $Y_L=0.1$. This is precisely the behavior found for the other nucleonic models in \cref{sec:deleptonization}. Indeed, as the leptons leave the core, the matter becomes more isospin asymmetric, and the symmetry energy contribution stiffens the EoS. The effective mass modifies this trend only mildly, the increase being slightly larger for the stiffer models with lower $m^*$.

The dependence on the entropy is weaker and, contrary to the deleptonization, its strength depends on the effective mass. For $m^*=0.75$ the maximum mass increases by $\approx 0.04\,M_\odot$ when the entropy is raised from $S/A=0$ to $S/A=2$, while the increase becomes progressively smaller for lower effective masses and, for $m^*=0.55$, $M_{\max}$ is practically insensitive to the entropy. Hence, the thermal pressure has a visible effect on $M_{\max}$ only for the softer, high-$m^*$ parametrizations. In all cases, the change of $M_{\max}$ produced by the cooling is smaller than the one produced by the deleptonization.

\begin{figure*}[tp]
    \centering
    \includegraphics[width=0.45\textwidth]{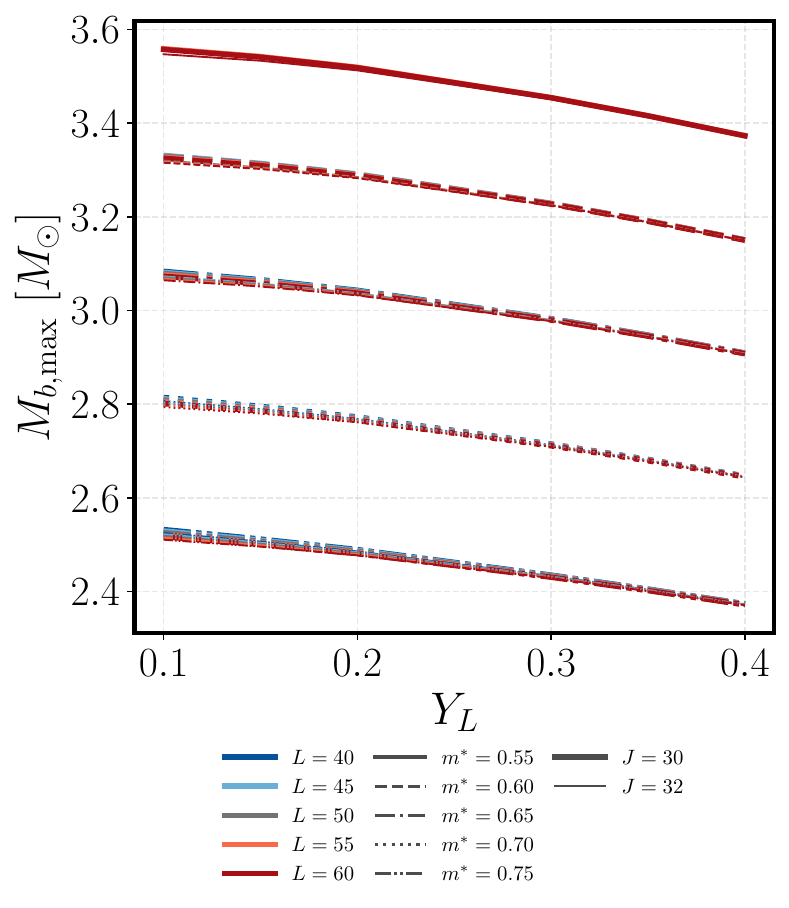}
    \hfill
    \includegraphics[width=0.45\textwidth]{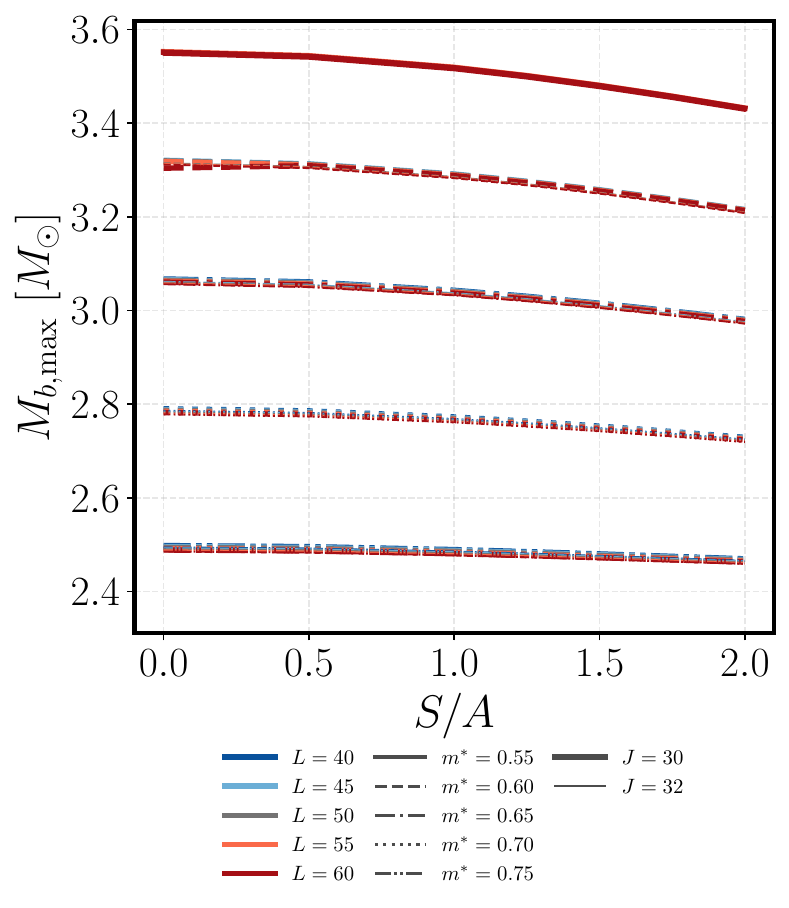}
    \caption{Same as \cref{fig:xeft_Mmax}, but for the maximum baryonic
    mass $M_{b,\max}$.}
    \label{fig:xeft_Mbmax}
\end{figure*}

The maximum baryonic mass is shown in \cref{fig:xeft_Mbmax}. As a function of the lepton fraction at $S/A=1$ (left panel), $M_{b,\max}$ follows the gravitational mass and increases monotonically with decreasing $Y_L$ for all parametrizations, by $0.14$--$0.19\,M_\odot$ between $Y_L=0.4$ and $Y_L=0.1$, with the larger values corresponding to the lower effective masses. The deleptonization therefore always enlarges the range of baryonic masses that can be supported: a star that is stable in the lepton-rich stage remains stable when its lepton abundance is reduced.

The entropy dependence of the maximum baryonic mass at constant $Y_L=0.2$ (right panel) shows a behavior that differs from the one of the gravitational mass not only in magnitude, but also in sign. As one can observe, $M_{b,\max}$ decreases monotonically with increasing entropy, by $\approx 0.12\,M_\odot$ for $m^*=0.55$ and by only $\approx 0.03\,M_\odot$ for $m^*=0.75$ between $S/A=0$ and $S/A=2$. In other words, contrary to the gravitational mass, the baryonic mass is less sensitive to the entropy for higher values of $m^*$. This behavior arises because the comparison is made at fixed entropy per baryon and not at fixed temperature. In the degenerate limit, the entropy per baryon fixes the ratio between the temperature and the Fermi energy of the nucleons, so that matter with a lower effective mass, which has a larger Fermi energy, is hotter and stores a larger thermal energy per baryon at the same $S/A$. This additional energy reduces the binding energy of the star and, therefore, the baryonic mass that can be supported. Nevertheless, for every parametrization the thermal reduction of $M_{b,\max}$ remains smaller than the increase produced by the deleptonization, even for $m^*=0.55$, where the heating from $S/A=0$ to $S/A=2$ at fixed $Y_L=0.2$ lowers it by $\approx 0.12\,M_\odot$, while the deleptonization from $Y_L=0.4$ to $Y_L=0.1$ at fixed $S/A=1$ raises it by $\approx 0.18\,M_\odot$. As found for the other nucleonic models in \cref{sec:heating_cooling}, the deleptonization therefore always raises the maximum baryonic mass, and by more than the heating lowers it.

\begin{figure*}[tp]
    \centering
    \includegraphics[width=\textwidth]{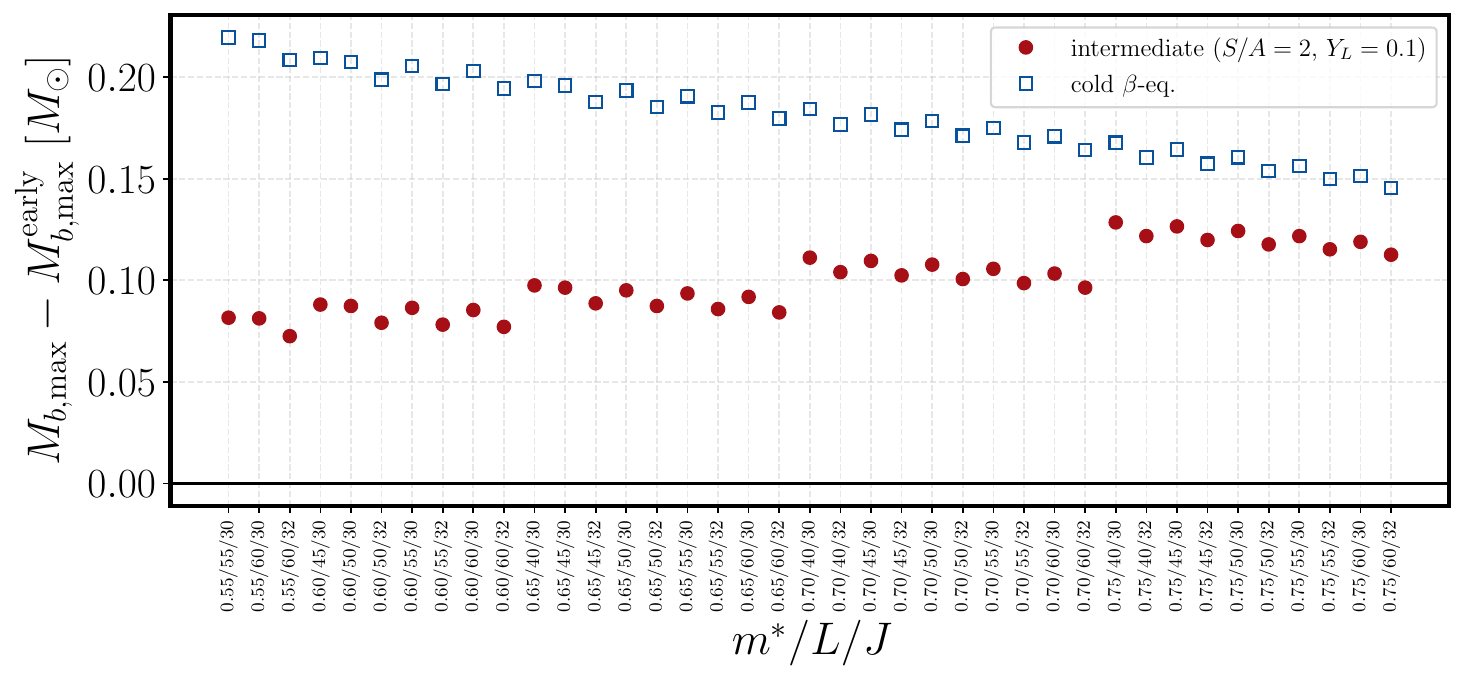}
    \caption{Change of the maximum baryonic mass along the PNS evolution for
    the $\chi$EFT-constrained RMF parametrizations, labeled on the horizontal
    axis as $m^*/L/J$ ($L$ and $J$ in MeV). Filled red circles: difference between
    the intermediate stage ($S/A=2$, $Y_L=0.1$) and the early stage
    ($S/A=1$, $Y_L=0.4$). Empty blue squares: difference between the cold,
    $\beta$-equilibrated star and the early stage. Positive values indicate
    that a star born with the maximum baryonic mass of the early stage
    remains stable at the later stage.}
    \label{fig:xeft_constant_Mb}
\end{figure*}

We finally address how the parameters of the model modify the stability windows between the three snapshots. In \cref{fig:xeft_constant_Mb} we therefore show, for the most massive stars, the difference between the maximum baryonic mass of the intermediate stage ($S/A=2$, $Y_L=0.1$) and that of the early stage ($S/A=1$, $Y_L=0.4$) (filled red circles), as well as the difference between the maximum baryonic mass of the cold, $\beta$-equilibrated star and that of the early stage (empty blue squares). Since the maximum baryonic mass of the early stage is the largest baryonic mass that a PNS can have at birth, a negative value of either difference would signal a delayed collapse along the constant-$M_b$ path. As can be seen, both differences are positive for all 39 parametrizations: the smallest margin, $\approx 0.07\,M_\odot$, is found for the intermediate stage of the stiffest ($m^*=0.55$) models, and the cold stars can always support between $0.15$ and $0.22\,M_\odot$ more baryonic mass than the early PNS. Moreover, the blue squares always lie above the red circles, so that for this nucleonic family the cold star can also support more baryonic mass than the intermediate stage. This is, however, not a general property of nucleonic matter, since for IUF, LPB, RB5, SRO(SLy4), STOS(TM1) and TM1e the maximum baryonic mass of the cold star is smaller than that of the intermediate stage, although still larger than that of the early PNS (see \cref{table:DelayedCollapse}). None of the RMF parametrizations predicts a delayed collapse of the most massive stars, in agreement with the other nucleonic models of \cref{table:DelayedCollapse}. It is worth noting that the BCPN and BCPY models of \cref{table:DelayedCollapse} are built on the same RMF Lagrangian~\cite{HornickRelativisticParametrizationNeutronMatterNS2018, BarmanInvestigatingRoleNuclearParametersOscillationModesHotNS2025}. While the purely nucleonic BCPN behaves as the parametrizations shown in \cref{fig:xeft_constant_Mb}, the inclusion of hyperons in BCPY reverses the sign of both mass differences, since the maximum baryonic mass of the intermediate and of the cold stage falls below that of the early stage by $0.128$ and $0.064\,M_\odot$, respectively, leading to a delayed collapse.

The size of the stability windows nevertheless follows a pattern, again set by the effective mass. The difference in $M_{b,\max}$ between the early and the intermediate stage increases with $m^*$, from typical values around $0.08\,M_\odot$ for $m^*=0.55$ to around $0.12\,M_\odot$ for $m^*=0.75$. Conversely, the difference in the baryonic mass between the intermediate stage and the cold star decreases with $m^*$, from $\approx 0.14\,M_\odot$ to $\approx 0.04\,M_\odot$, and so does the difference in baryonic mass between the early stage and the cold star. This pattern can be understood from the results discussed above. The transition from the early to the intermediate stage combines a deleptonization, which increases $M_{b,\max}$, with a heating from $S/A=1$ to $S/A=2$, which decreases it. Since the latter effect is much stronger for low effective masses (see the right panel of Fig.~\ref{fig:xeft_Mbmax}), the net gain is reduced for the stiffest parametrizations. In the subsequent transition to the cold star, the situation is reversed. Indeed, as the star cools down from $S/A=2$ to a cold configuration, the binding energy of the star is increased, so the models with the strongest entropy dependence of $M_{b,\max}$, i.e.~those with low $m^*$, experience the largest gain.

We conclude that, within a family of nucleonic RMF models that are consistent with the zero-temperature $\chi$EFT constraints on neutron matter, and that cover a wide range of stiffness and of the symmetry energy parameters, no parametrization predicts a delayed collapse of the most massive PNS. For all these parametrizations, both the deleptonization and the cooling increase the maximum baryonic mass that can be supported, and the parameters of the model only modify the size of the stability windows without switching to an instability. These results support the conclusion that the delayed collapse is not a feature that can be easily obtained by varying the parameters of a realistic nucleonic EoS, but rather an effect triggered by the appearance of exotic hadrons, which soften the EoS. A systematic scan of the hyperonic sector using chiral EFT is not feasible, since the hyperon-nucleon and hyperon-hyperon interactions are not
constrained at the high densities where hyperons appear. The equivalent exercise would be a scan of the hyperonic coupling ratios within the ranges allowed by SU(3) symmetry and by hypernuclear data~\cite{Weissenborn:2011kb, Weissenborn:2011ut, Fortin:2017cvt}, which we leave for future work.
\section{Summary and conclusions}
\label{sec:conclusions}
 
In this work, we have revisited the evolution of the massive proto-neutron stars (PNSs) in the presence of additional hadronic degrees of freedom and compared it to the case when only nucleons are present. The analysis aimed to assess whether the delayed collapse of the massive PNSs can indeed be regarded as an indicator of the appearance of additional hadrons besides nucleons in the dense core of the star. 
We employed 39 general-purpose finite-temperature EoSs  
available on the CompOSE database~\cite{TypelCompOSE2015, TypelCompOSEManual2022}, 16 of which contain additional hadronic degrees of freedom. The selection of models is intentionally heterogeneous, as it covers non-linear and density-dependent covariant density functionals, the chiral mean-field models, the quark-meson-coupling schemes, non-relativistic Skyrme interactions, and microscopic approaches, so that the conclusions drawn here do not rely on the particularities of a single model or framework.
 
Besides the instability along the constant baryonic mass path, we have studied other quantities that characterize the maximum-mass configuration, namely the maximum gravitational and baryonic masses, the central density, the radius and the effective thermal index of matter. To properly disentangle the role played by the different physical processes at work during the early evolution of NSs, we have studied the deleptonization of the matter (at fixed entropy per baryon) and its heating and cooling (at fixed lepton fraction). In addition, we have strengthened the conclusions obtained for purely nucleonic matter by performing a systematic analysis within the RMF framework, employing parametrizations that satisfy the constraints coming from chiral effective field theory~\cite{HornickRelativisticParametrizationNeutronMatterNS2018, Christian:2023byx}.
 
Our main findings are summarized in the following points.
 
\begin{enumerate}
 
\item We have confirmed that there is an instability for massive PNSs at constant baryonic mass, which can be traced back to the appearance of hyperons in the stellar core. Out of the 15 EoSs in our sample that contain hyperons, 13 predict such an instability, while none of the 23 purely nucleonic models do. This behavior is not an artifact of the particular models employed. The softening that drives the instability has a largely thermodynamic origin, since the hyperons that populate the medium during the deleptonization open new Fermi seas and thereby reduce the degeneracy pressure of the nucleons and leptons. The details of the hyperonic interaction only determine the onset and the abundance of the hyperons, and hence the size of the effect, but not its presence. Therefore, if hyperons appear in sizable amounts in the core of the most massive PNSs, the delayed collapse is unavoidable.

In the purely nucleonic models, on the contrary, the EoS becomes stiffer as the matter is driven towards larger isospin asymmetries by the escaping neutrinos.

 \item The central density of the maximum-mass configuration obtained with most of the hyperonic models increases during the deleptonization, while for the nucleonic models it does not change significantly. The maximum-mass star contracts in both cases, by up to about $1.6$~km for the hyperonic models and at most $0.9$~km for the nucleonic ones, but the two families differ in which part of the star contracts. The outer layers contract by $0.5$ to $1.2$~km irrespective of composition, whereas the core shrinks only in the hyperonic case, by $0.1$ to $0.6$~km, and expands for the nucleonic models. The contraction of the core due to the presence of hyperons can lead, e.g., to an increased emissivity of neutrinos due to the increased core density and the opening of the hyperon direct Urca process, whereas both effects would be absent for the purely nucleonic case.
 
\item Although the hyperonic EoSs are known to display a distinct thermal behavior~\cite{KochankovskiEoSHotHyperonicNSM2022, RadutaEoSHotNSExoticParticleDoF2022, BlackerThermalBehaviorIndicatorHyperonsBNSMergers2024}, the thermal effects play a sub-leading role in the evolution of the most massive PNSs, since they do not by themselves discriminate between the two classes of models. The heating lowers the maximum baryonic mass of most of the EoSs, both hyperonic and nucleonic. The reduction is slightly larger in the presence of hyperons. However, it is the opposite effect due to deleptonization that decides on the different fate, collapse or no collapse, of the two families. Similarly, the effective thermal index decreases under heating for most of the hyperonic models, and increases for most of the nucleonic ones, but the nucleonic models with a soft thermal response, i.e.\ an effective thermal index $\Gamma_{\rm th}$ that decreases under heating (LPB and the Skyrme models RB2, RB5, RB7), display the same decreasing trend.  
A purely nucleonic model therefore cannot mimic the delayed collapse by being thermally soft alone. The thermal effects are not irrelevant, however. In the QMC-A model, where hyperons appear only at high densities, the deleptonization does not increase the maximum baryonic mass, so that the thermal loss is not compensated and the instability appears (only) at the hot intermediate stage. Its cold maximum baryonic mass is slightly larger than that at the early stage of PNS evolution, so it is the heating to the intermediate stage that triggers the collapse. The models BHB$\Lambda\phi$ and DD2Y$\Delta$(1.2,1.1) also predict instabilities between the first two snapshots of PNS evolution. The hyperonic models SFHoY and DNS suffer a comparable thermal loss, but there the deleptonization still compensates it, and these models do not collapse.
 
\item While both hyperons and kaons soften the EoS and can induce a delayed collapse, the composition of stars containing either hyperons or an antikaon condensate evolves differently. For most of the hyperonic models, the mass-averaged hyperon fraction of the maximum-mass star peaks in the hot, partially deleptonized stage and is reduced as the star cools. The kaon abundance, on the contrary, grows monotonically and attains its largest value in the cold star. Both species are favored by the deleptonization. The hyperons are thermally populated below their cold threshold and therefore peak in the hot intermediate stage, whereas the condensate, being a zero-momentum state, is suppressed by the temperature and keeps growing as the star cools. Even a kaonic fraction small compared with the typical hyperonic ones is sufficient to trigger the instability, as illustrated by DD2K, which already collapses at the intermediate stage, where $\langle Y_K \rangle \approx 0.08$.
 
\item The exotic components also leave imprints on the thermal and structural properties of the maximum-mass star. At fixed entropy per baryon, the entropy is shared among a larger number of particle species, so that hyperon- and resonance-rich stars are colder than their nucleonic counterparts. For models sharing the same nucleonic interaction, the mass-averaged temperature is lowered by 16\% to 22\% in the early stage and by 22\% to 30\% in the intermediate one, and by up to 35\% when $\Delta$ resonances are present. For the same reason, the deleptonization at fixed entropy per baryon heats the nucleonic stars by mean temperatures of about 6 to 8~MeV and the hyperonic ones by only 1 to 2~MeV. Moreover, between the early and the intermediate stage the hyperon-rich stars contract and become denser, whereas the nucleonic ones expand, so that the models separate according to their hyperon content in the plane of the change of central density against the change of radius.
 
\item Finally, the systematic exploration of the couplings within a single RMF model, restricted to the parametrizations that remain compatible with the chiral effective field theory constraints on pure neutron matter, shows that the response of purely nucleonic matter to the deleptonization does not depend on the particular choice of parameters. In all the cases considered, the deleptonization raises the maximum baryonic mass by more than the heating lowers it, so that no instability along constant baryonic mass is encountered. The parameters only modify the size of these stability windows, not their presence, which supports the interpretation of the instability as a signature of additional hadronic degrees of freedom.
 
\end{enumerate}
 
The delayed collapse scenario and the role played in it by the deleptonization were already pointed out in early works~\cite{BombaciMmaxNS1996, PrakashCompositionStructurePNS1997}, but a systematic confirmation covering a large and diverse set of models was still lacking, and it is this gap that the present study intends to fill. Nevertheless, a few words of caution are in order. First, all the finite-temperature hyperonic EoSs that are available today, and therefore all those employed here, are phenomenological. Since the mechanism behind the instability is largely statistical, this limitation affects mainly the onset and the abundance of the hyperons, and hence the size of the metastability window, rather than the qualitative picture. Future works should try to address this issue by extending the analysis to EoSs obtained either with ab initio many-body methods~\cite{SchulzeYNMatterBHF1998, VidanaStrangeNuclearMatterBHF2000, BaldoYStarsBBG2000, BurgioHyperonStarsFiniteTBruecknerTheory2011} or with more complete density functional approaches~\cite{KochankovskiMomentumFragmentedHyperonsColdNSs2026}. Second, as discussed in \cref{sec:PNSevolution}, our snapshots assume constant lepton fraction and entropy per baryon throughout the star and neglect dynamical features such as rotation~\cite{daSilvaHyperonsDeltaResonancesRotatingPNS2025} and the possibility that some of the weak reactions are slow compared with the dynamical timescale~\cite{ZatiniLambdaHyperonsCCSNEquilibrium2026, CanullanPascualNSStabilityBeyondMassPeakOutEquilibriumPerturbations2025}. These refinements are expected to shift the precise boundaries of the metastability window, but not to alter the qualitative separation between the two families of models that we have established here. 
 
The natural continuation of this work is to connect the systematics presented above with the observable signatures of the delayed collapse, namely the neutrino signal and the spin-up of the star. The most direct one is the abrupt cessation of the neutrino signal, which was shown long ago to distinguish a metastable from a stable evolutionary track~\cite{PonsEvolutionPNSKaonCondensates2001, PonsEvolutionPNS1999}, and which has recently been revisited with detailed Boltzmann neutrino transport~\cite{FischerHyperonsPNSDeleptonizationDarkFlavouredParticles2025}. A second one is the distinctive spin-up of the hyperonic PNS that precedes the collapse~\cite{YuanRotationalEvolutionPNSHyperons2005}. Performing full Kelvin-Helmholtz evolutions with the models identified here as metastable, and in particular with the controlled sets of EoSs that share the same nucleonic interaction, would allow one to quantify how the difference in the compositional evolution translates into the timing and the shape of these signals. 
 
\section*{Acknowledgments}
This work is supported by the grants CEX2020-001058-M and CEX2024-001451-M of the Spanish program ``Unidad de Excelencia María de Maeztu'', financed by MCIN/AEI/10.13039/501100011033, the MaX-CSIC Excellence Award MaX4-SOMMA-ICE, the projects PID2023-147112NB-C21, PID2022-139427NB-I00 and PID2025-168786NB-I00 financed by the Spanish MCIN/AEI/10.13039/501100011033 and ERDF/EU, and by the Grant CIPROM 2023/59 of Generalitat Valenciana. I. A. R., J. S. B., and L. T. acknowledge support from the Deutsche Forschungsgemeinschaft (DFG, German Research Foundation) through the CRC-TR 211 'Strong-interaction matter under extreme conditions' -- project number 315477589 -- TRR 211. I. A. R. also acknowledges support from the Deutsche Forschungsgemeinschaft (DFG, German Research Foundation) – Project Number 579861443. J. E. C. was supported by the European Research Council (ERC) Advanced Grant INSPIRATION under the European Union's Horizon 2020 research and innovation programme (Grant agreement No. 101053985).
\makeatletter\if@filesw\immediate\write\@auxout{\string\citation{apsrevControlHidePublishedEprints}}\fi\makeatother % hide arXiv numbers of published articles (control entry in bibliography.bib)
\bibliography{bibliography}

\end{document}